\documentclass[twocolumn]{aastex61}
\pdfoutput=1
\usepackage{amsmath,amstext}
\usepackage[T1]{fontenc}
\usepackage[figure,figure*]{hypcap}
\usepackage{tikz}
\usetikzlibrary{shapes,arrows}

\shorttitle{Analysis of DA and DB white dwarfs from SDSS}
\shortauthors{Genest-Beaulieu \& Bergeron}

\begin{document}

\title{A comprehensive spectroscopic and photometric analysis of DA and DB white 
dwarfs from SDSS and {\it Gaia}}

\author{C. Genest-Beaulieu}
\affiliation{D\'epartement de Physique, Universit\'e de Montr\'eal, Montr\'eal, 
QC H3C 3J7, Canada; genest@astro.umontreal.ca, bergeron@astro.umontreal.ca.}
\author{P. Bergeron}
\affiliation{D\'epartement de Physique, Universit\'e de Montr\'eal, Montr\'eal, 
QC H3C 3J7, Canada; genest@astro.umontreal.ca, bergeron@astro.umontreal.ca.}

\begin{abstract}
We present a detailed spectroscopic and photometric analysis of DA and
DB white dwarfs drawn from the Sloan Digital Sky Survey with
trigonometric parallax measurements available from the {\it Gaia}
mission. The temperature and mass scales obtained from fits to $ugriz$
photometry appear reasonable for both DA and DB stars, with almost
identical mean masses of $\langle M \rangle = 0.617~M_\odot$ and
$0.620~M_\odot$, respectively. The comparison with similar results
obtained from spectroscopy reveals several problems with our model
spectra for both pure hydrogen and pure helium compositions. In
particular, we find that the spectroscopic temperatures of DA stars
exceed the photometric values by $\sim$10\% above $T_{\rm
  eff}\sim14,000$~K, while for DB white dwarfs, we observe large
differences between photometric and spectroscopic masses below $T_{\rm
  eff}\sim16,000$~K. We attribute these discrepancies to the
inaccurate treatment of Stark and van der Waals broadening in our
model spectra, respectively. Despite these problems, the mean masses
derived from spectroscopy --- $\langle M \rangle = 0.615~M_\odot$ and
$0.625~M_\odot$ for the DA and DB stars, respectively --- agree
extremely well with those obtained from photometry. Our analysis also
reveals the presence of several unresolved double degenerate binaries,
including DA+DA, DB+DB, DA+DB, and even DA+DC systems. We finally take
advantage of the {\it Gaia} parallaxes to test the theoretical
mass-radius relation for white dwarfs. We find that 65\% of the white
dwarfs are consistent within the 1$\sigma$ confidence level with the
predictions of the mass-radius relation, thus providing strong support
to the theory of stellar degeneracy.

\end{abstract}


\section{Introduction}

Our understanding of white dwarf stars relies heavily on the
determination of their physical parameters, such as effective
temperature, surface gravity, luminosity, mass, radius, atmospheric
composition, and cooling age. Several independent methods have been
developed over the years to measure directly some of these parameters,
while others are obtained indirectly through detailed evolutionary
models.

The most widely used method, at least until now, to measure the
effective temperature ($T_{\rm eff}$) and the surface gravity ($\log
g$) of white dwarfs involves the comparison of the observed and model
spectra, known as the spectroscopic technique. First applied to a
large sample of DA stars by \cite{BSL1992}, this technique has been
used repeatedly since then in several other studies (see, e.g.,
\citealt{LBH05}, \citealt{Koester2009SDSS}, \citealt{Koester2009},
\citealt{Tremblay2011}, \citealt{Gianninas2011},
\citealt{Genest2014}). A similar approach has also been applied in the
context of DB white dwarfs (\citealt{Eisenstein2006},
\citealt{Voss2007}, \citealt{Bergeron2011}, \citealt{Koester2015},
\citealt{Rolland2018}); in this case, the hydrogen abundance is also
being measured spectroscopically.

Another method that can be applied to large ensembles of white dwarfs
is the photometric technique, where the observed energy distribution,
built from magnitudes in various bandpasses, is compared to the
predictions from model atmospheres \citep[see,
  e.g.,][]{Bergeron1997}. This method yields the effective temperature
and the solid angle $\pi (R/D)^2$ of the star; if the trigonometric
parallax (or distance) is known, the radius can be obtained directly.
Until recently, trigonometric parallaxes were available for only a few
hundreds, mostly cool white dwarfs (\citealt{Bergeron2001},
\citealt{Holberg12}, \citealt{Tremblay17}, \citealt{Bedard2017}). In
the absence of trigonometric parallax measurements, one usually
assumes a value of $\log g=8$, in which case the photometric method
can be applied to large white dwarf samples, such as those identified
in the Sloan Digital Sky Survey (SDSS, \citealt{Genest2014}). In some
situations, the photometric technique is the only method applicable,
for example for cool white dwarfs, which present no absorption
features.

With either the spectroscopic or photometric techniques, the mass of
the white dwarf can only be obtained from detailed evolutionary
models, which provide the required temperature-dependent relation
between the mass and the radius, as well as cooling ages. This
theoretical mass-radius relation for white dwarfs has recently been
tested, but only for relatively small samples (\citealt{Tremblay17},
\citealt{Parsons17}, \citealt{Bedard2017}), a situation that is about
to change by taking advantage of the recently measured trigonometric
parallaxes from {\it Gaia} \citep{gaia2}.

One of the most important issue regarding both the spectroscopic and
photometric techniques is the precision and accuracy of each method.
Statistically speaking, the precision of the method describes random
errors, a measure of statistical variability, repeatability, or
reproducibility of the measurement, while the accuracy represents the
proximity of the measurements to the true value being measured, in our
case, the true $T_{\rm eff}$ and $\log g$ (or mass) values. It has
been argued repeatedly in the literature that the spectroscopic
technique yields more precise atmospheric parameters than the
photometric technique, in general because of the moderate quality of
photometric and parallax measurements. However, the exquisite parallax
data from {\it Gaia} and photometric data from SDSS or Pan-STARRS may
change this picture drastically in favor of the photometric approach.

Another advantage of the photometric technique is that the synthetic
photometry is less sensitive to the input physics included in the
model atmospheres. Indeed, the shape and strength of the spectral
lines are affected by a number of factors (line broadening, convective
energy transport, etc.), which in turn affect the atmospheric
parameters measured spectroscopically. One well-known example is the
so-called high-$\log g$ problem in cool DA white dwarfs, which was
explained by an inadequate treatment of convection in standard 1D model
atmospheres \citep{Tremblay2013}.

The question of the precision and accuracy of the spectroscopic and
photometric methods using various white dwarf samples and photometric
data sets has been discussed at length by \citet[][see also
  \citealt{Gentile2019}]{Tremblay2019}. Because of the utmost
importance of this issue for the white dwarf field, and also because
of our different approach to the problem, we present in this paper our
own independent assessment of the internal consistency between both
fitting techniques by comparing the atmospheric and physical
parameters of DA and DB white dwarfs obtained from spectroscopy with
those derived from the photometric technique. The white dwarf samples
used in this study are presented in Section \ref{sect:sample},
followed by a brief description of our models in Section
\ref{sect:theory}.  Our photometric and spectroscopic analyses are
presented in Sections \ref{sect:Photometry} and
\ref{sect:Spectroscopy}, respectively. Section \ref{sect:comparison}
is dedicated to the comparison of the atmospheric parameters obtained
from photometry and spectroscopy. Using these results, the theoretical
mass-radius relation is then put to the test in Section
\ref{sect:MR}. We conclude in Section \ref{sect:conclusion}.

\section{Sample}\label{sect:sample}

The primary goal of this study is to compare the atmospheric
parameters of DA and DB white dwarfs obtained from different
techniques and data types. Given that the Sloan Digital Sky Survey
(SDSS) provides photometry and spectroscopy for over 30,000 white
dwarfs, we base our analysis on this particular data set. We started
by retrieving all spectroscopic and photometric data for all DA and DB
white dwarfs --- including all subtypes (DAH, DB:, DBZ, etc.) ---
spectroscopically identified in the SDSS, up to the Data Release 12
\citep{DR7,DR10,DR12}. This represents a total of 27,217 DA and 2227
DB spectra, with corresponding $ugriz$ photometric data sets. We also
want to take advantage of the {\it Gaia} DR2 catalog \citep{gaia2},
which contains precise trigonometric parallax measurements for a large
number of SDSS objects. In order to ensure that the atmospheric
parameters we derive are reliable, we need to apply a few criteria to
keep only the best SDSS and {\it Gaia} data sets.

\begin{figure}[t]
\includegraphics[width=\linewidth]{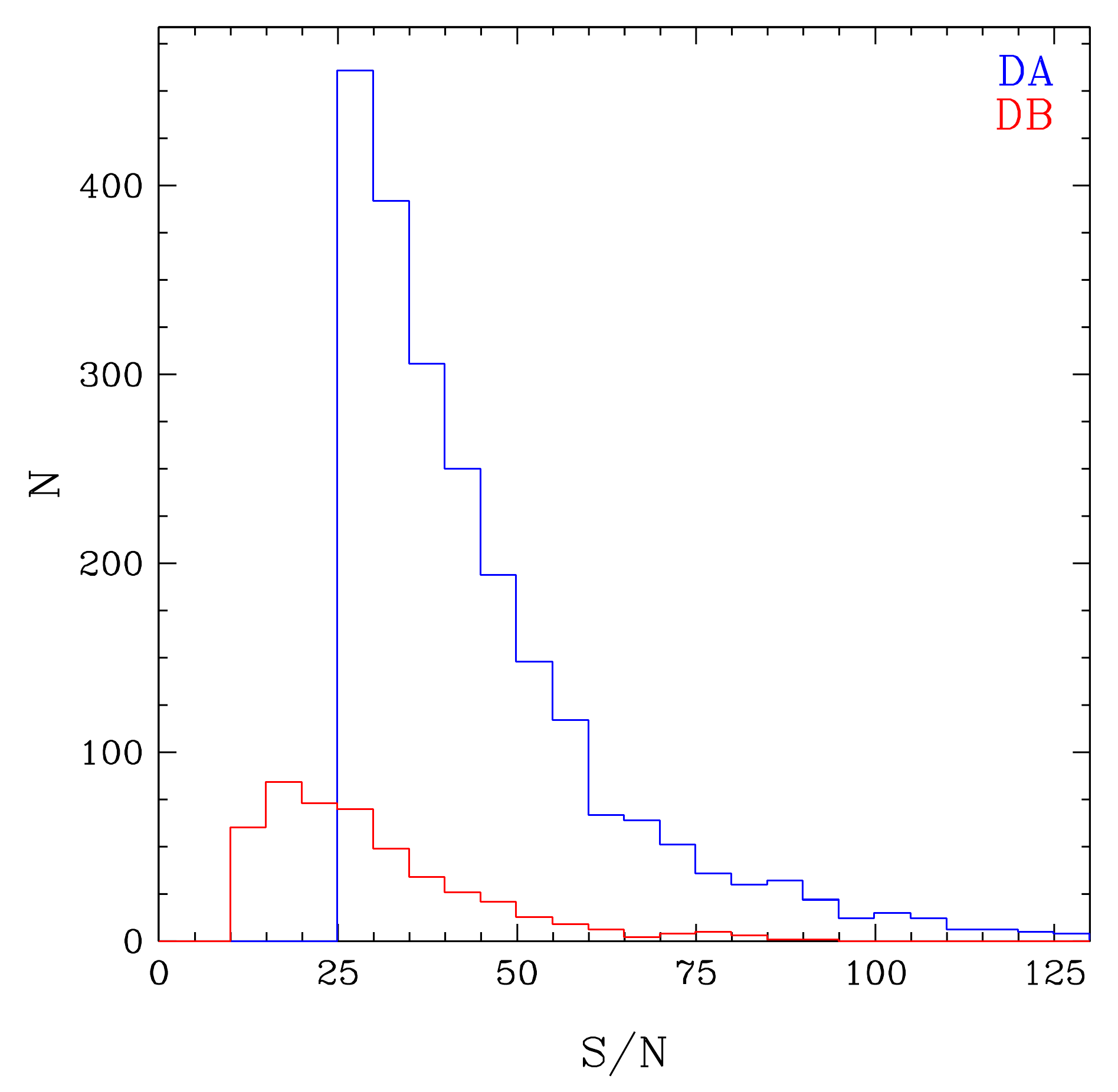}
\caption{Distribution of signal-to-noise ratios of the DA (blue)
  and DB (red) white dwarfs in our sample.}
\label{fig:SN}
\end{figure}

We first removed every object with a spectral type indicating a
magnetic object (H), a known companion (+ and/or M), emission lines
(E), or an uncertain spectral type (:). For the DA sample, we also
removed any spectral type indicating the presence of helium (B or O)
or metals (Z). Therefore, our sample contains only the spectral types
DA, DB, DBA(Z), and DBZ(A). We then applied a lower limit on the
signal-to-noise ratio (S/N) of the SDSS optical spectra. Given the
very large number of DA white dwarfs, we chose to keep only those with
${\rm S/N} \geq 25$. For the DB white dwarfs, which are not as common
as DA stars, we chose to set the limit at a lower value of ${\rm S/N}
\geq 10$. The S/N distribution of spectra in our sample is displayed
in Figure \ref{fig:SN}. Finally, we kept only the objects with {\it Gaia}
parallax measurements more precise than 10\% ($\sigma_\pi/\pi \leq 0.1$).

After applying all these criteria, we are left with 2236 and 461
individual spectra and corresponding $ugriz$ data sets for DA and DB
white dwarfs, respectively.  Since the calibration algorithm has
changed between DR7 and DR8, and that the $ugriz$ zeropoints have been
recalibrated in
DR13\footnote{https://www.sdss.org/dr14/algorithms/fluxcal/}, we use
the $ugriz$ magnitudes from the SDSS DR14 instead of the values given
in the aforementioned catalogs.

\begin{figure}[t]
\includegraphics[width=\linewidth]{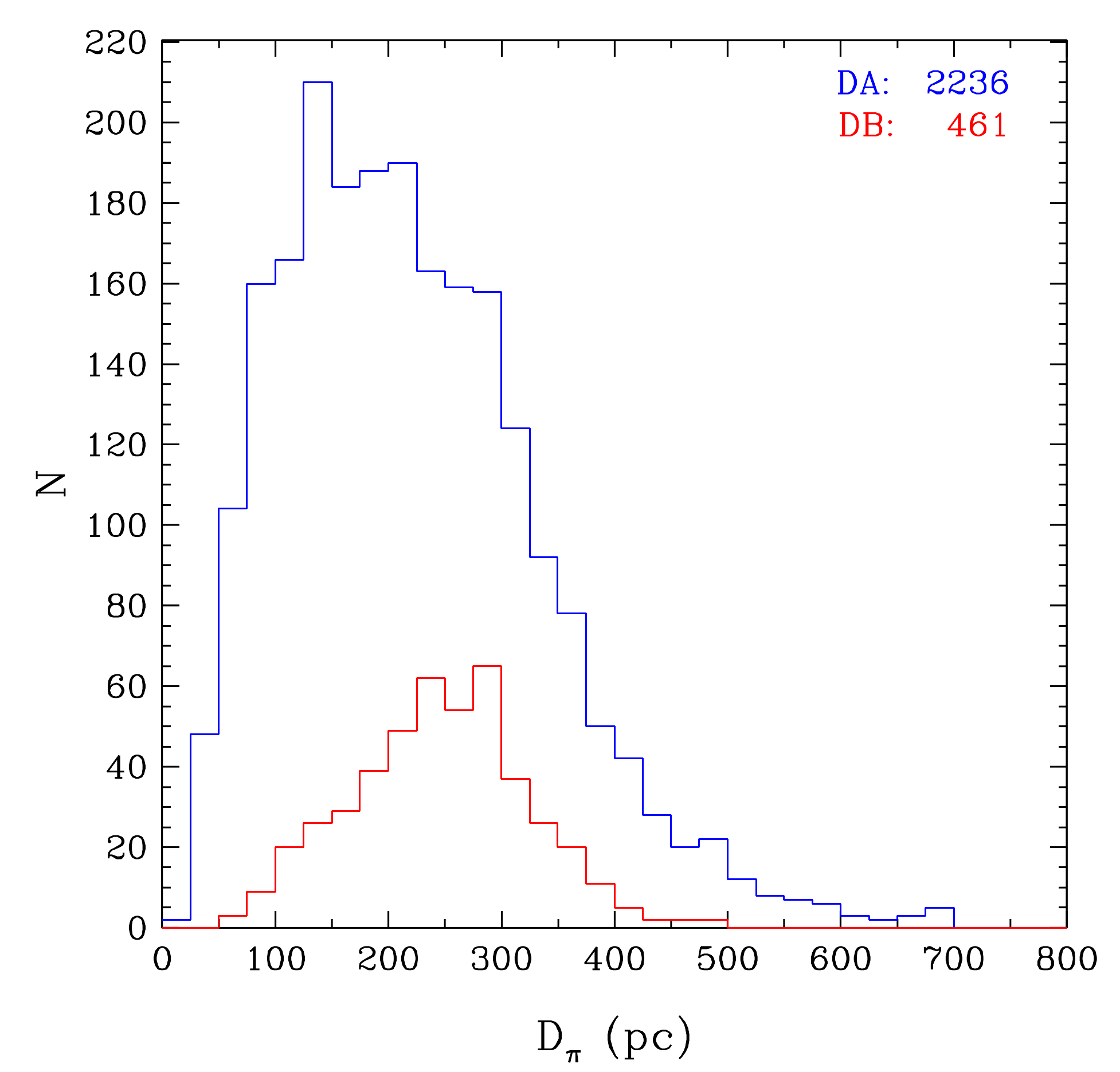}
\caption{Distribution of parallactic distances for the DA (blue)
  and DB (red) white dwarfs in our sample.}
\label{fig:distances}
\end{figure}

Figure \ref{fig:distances} presents the distribution of the white
dwarfs in our sample as a function of distance and spectral type. We
note that most of the objects in our sample are located at large
distances where interstellar reddening becomes important ($D\gtrsim 100$
pc). This will be discussed in section \ref{sect:techphot}.

\section{Theoretical Framework}\label{sect:theory}

Since our sample contains both DA and DB white dwarfs, we require two
different grids of model atmospheres and synthetic spectra.

\subsection{DA Model Atmospheres}\label{sect:modelDA}

Our pure hydrogen grid for the DA stars is calculated with two
different codes. For $T_{\rm eff} < 30,000~{\rm K}$, we use the LTE
version described at length in \cite{Tremblay2009}. Convective energy
transport, which becomes important below $T_{\rm eff} \sim 15,000~{\rm
  K}$, is treated with the ML2/$\alpha=0.7$ version of the
mixing-length theory (MLT). The adopted parameterization of the MLT is
important, particularly for cool DA white dwarfs, and it affects
mostly the atmospheric parameters determined from spectroscopic data
(see section \ref{sect:specresults}). Above 30,000 K, NLTE effects are
taken into account using TLUSTY \citep{Hubeny1995}. Combining both
grids, we obtain model spectra ranging from $T_{\rm eff} = 1500~{\rm
  K}$ up to $120,000~{\rm K}$, with surface gravities between $\log g
= 6.5$ and $9.0$. Note that both model grids rely on the improved
Stark profiles of \cite{Tremblay2009}.

\subsection{DB/DBA Model Atmospheres}\label{sect:modelDB}

Our DB/DBA model grid is similar to that described in
\cite{Bergeron2011} (but see section \ref{sect:ivdw}). These models
are in LTE and calculated using the ML2/$\alpha=1.25$ parameterization
of the MLT.  Our grid covers effective temperatures between 11,000 K
and 50,000 K, surface gravities ranging from $\log g = 7.0$ to $9.0$,
and hydrogen abundances from $\log N({\rm H})/N({\rm He})=-6.5$ to
$-2.0$, and an additional pure helium grid.
 
Below $T_{\rm eff}\sim16,000$~K, He~\textsc{i} line broadening by
neutral particles becomes important. We can divide this into two
parts: van der Waals and resonance broadening. In our models, these
two broadening mechanisms are combined following the procedure
described in detail by \cite{BeauchampPhD}, which we summarize
here. Resonance broadening is treated according to the theory of
\cite{Ali1965}, while van der Waals broadening requires a more
elaborate approach.  The width of the Lorentzian profile ($\omega_{\rm
  vdW}$) is calculated twice, once with the theory described in
\cite{Unsold1955}, and the second time by following the approach
discussed in \cite{Deridder1976}. The Deridder \& van Rensbergen
theory systematically predicts a larger profile if the initial and
final effective quantum numbers of the transition are higher than 2 or
3. Since they used the Smirnov potential, which becomes invalid below
these numbers, we keep the following conservative value for the width
of the Lorentzian profile:

\begin{equation}
\omega_{\rm vdW}={\rm max}(\omega_{\rm Unsold},\omega_{\rm Deridder})\ .
\end{equation}
 
\noindent It was also found empirically by \cite{BeauchampPhD} that
the neutral helium lines of cool DB stars at $\lambda=4121$ \AA\ and
4713 \AA\ could be better reproduced it they were strictly treated
within the \cite{Unsold1955} theory, which is the procedure we adopt
here as well.
 
Finally, \cite{Lewis1967} found that the combination of the resonance and van
der Waals broadening led to a profile with $\omega_{\rm
  neutral}\sim0.6-0.8\,(\omega_{\rm resonance}+\omega_{\rm
  vdW})$. However, that study was only valid for temperatures around
100 K, hardly applicable to our DB models. Therefore, the conservative
value of 
 
\begin{equation}
\omega_{\rm neutral}={\rm max}(\omega_{\rm resonance},\omega_{\rm vdW})
\end{equation}
 
\noindent is adopted instead. For simplicity, we will refer to the above
procedure as the Deridder \& van Rensbergen theory.
   
Note that more recent self-broadening calculations of helium lines
have been performed by \citet{Leo1995}, but their results are only
available at temperatures of $T=80~{\rm K}$ and $300~{\rm K}$, hardly
applicable in the context of our DB models, as before.

\section{Photometric Analysis}\label{sect:Photometry}

The first step in our analysis is to measure the atmospheric and physical
parameters derived from photometry. Before presenting the results, we
first describe the technique used to determine these parameters, and
we also explore the effects of the presence of atmospheric hydrogen
on the photometric solutions of DB white dwarfs.

\subsection{Photometric Technique}\label{sect:techphot}

The photometric technique relies on the energy distribution to measure
the effective temperature and stellar radius. We use here the
method described at length in \cite{Bergeron1997}, which is the same for both
DA and DB stars. Since we are using the SDSS $ugriz$ photometry, we
first need to apply the corrections to the $u$, $i$, and $z$ bands to
account for the transformation from the SDSS to the AB magnitude
system. These corrections, given in \cite{Eisenstein2006}, are

\begin{equation}\label{eq:correctionsSDSS}
\begin{split}
u &= u_{\rm SDSS}-0.040 \\
i &= i_{\rm SDSS}+0.015 \\
z &= z_{\rm SDSS}+0.030
\end{split}
\end{equation}

\noindent For completeness, we repeated the experiment displayed in
Figure 8 of \citet{Genest2014} where observed magnitudes are compared
with those predicted by the photometric technique, and we found that
the above constants are still appropriate, and lead to a better
agreement between bots sets of magnitudes.

One important aspect to consider while dealing with photometric
observations is interstellar reddening, which becomes important for
$D\gtrsim100$ pc. As can be seen from Figure \ref{fig:distances}, the
majority of our objects are located beyond 100 pc, implying that
interstellar extinction cannot be neglected in our analysis. The
procedure used here is based on the approach described by
\citet{Harris2006}, where the extinction is assumed to be negligible
for stars with distances less than 100 pc, to be maximum for those
located at $|z|>250$ pc from the galactic plane, and to vary linearly
along the line of sight between these two regimes. We also explore in
Section \ref{sect:Teff} an alternative procedure proposed by
\citet{Gentile2019}. Since the trigonometric parallax is known for
every object in our sample, magnitudes can be dereddened
directly. This procedure is accomplished using the $E(B-V)$ values
from \cite{SF2011}.

Every dereddened magnitude $m_\nu$ is then converted into an average
flux $f^m_\nu$ using the relation

\begin{equation}
m_\nu = - 2.5 \log f^m_\nu - 48.60
\end{equation}

\noindent where

\begin{equation}\label{eq:fluxmoyen}
f^m_\nu=\frac{\int f_\nu S_m(\nu)\,d\log \nu}{\int S_m(\nu)\,d\log \nu}\ ,
\end{equation}

\noindent and where $f_\nu$ is the monochromatic flux from the star received at
Earth, and $S_m(\nu)$ is the total system response, including
atmospheric transmission and mirror reflectance.

\begin{figure}
\includegraphics[width=\columnwidth]{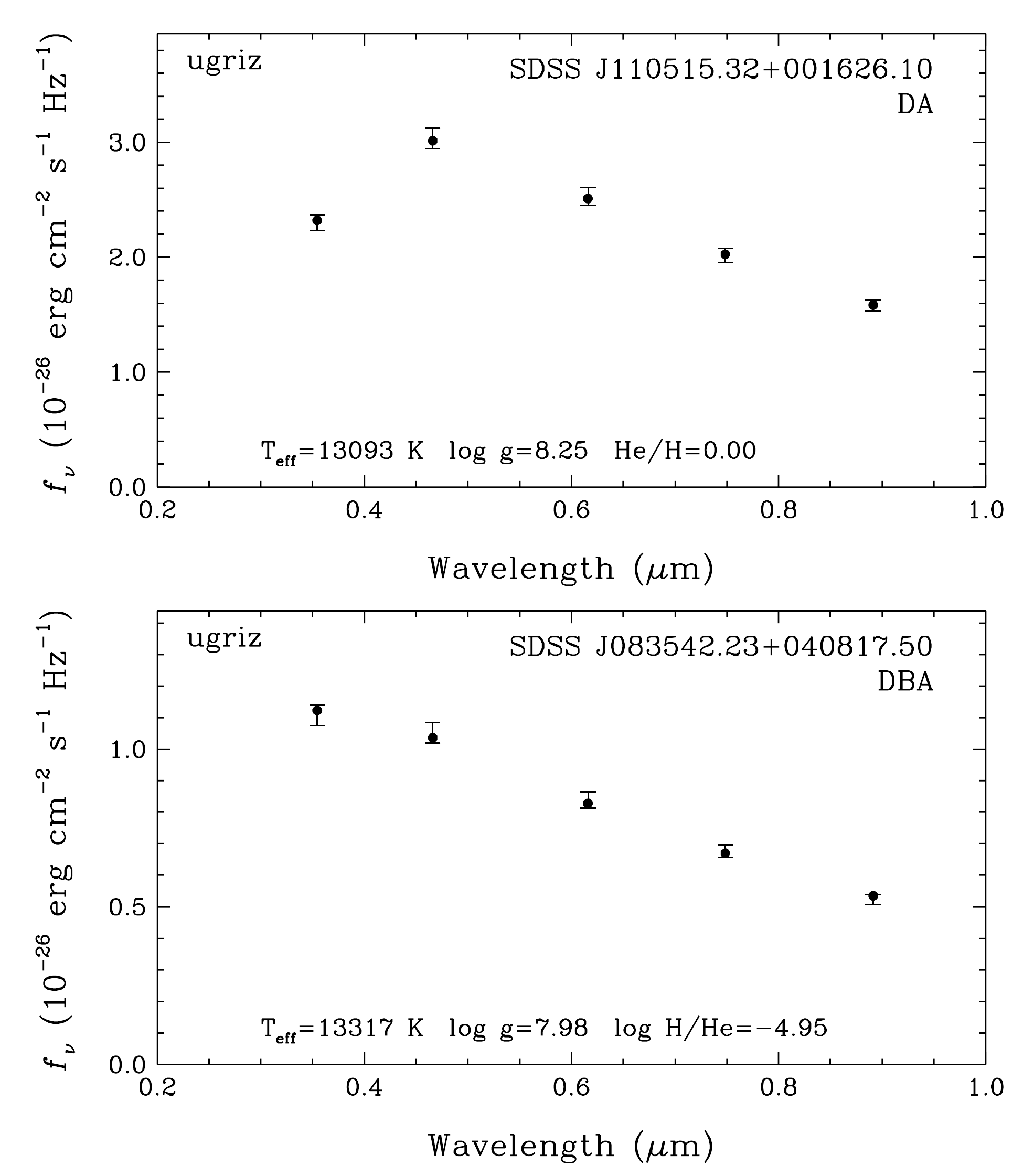}
\caption{Examples of the photometric technique for the DA white dwarf
  SDSS J110515.32+001626.10 (top) and the DBA SDSS
  J083542.23+040817.50 (bottom). The error bars represent the observed
  data, while the best-fit model is shown by the filled
  circles. The resulting atmospheric parameters are given in each
  panel.}
\label{fig:techphot}
\end{figure}

The same conversion is performed using our synthetic spectra. We
obtain the average synthetic fluxes, $H_\nu^m$, by substituting
$f_\nu$ in equation \ref{eq:fluxmoyen} with the monochromatic
Eddington flux $H_\nu$. The average observed and model fluxes are
related through the equation

\begin{equation}
f^m_\nu=4\pi\left(R/D\right)^2H_\nu^m
\end{equation} 

\noindent where $R$ is the radius of the white dwarf, and $D$ its
distance from Earth. We then proceed to minimize the $\chi^2$ value,
which is defined in terms of the difference between observed and model
fluxes over all bandpasses, properly weighted by the photometric
uncertainties. Our minimization procedure relies on the non-linear
least-squares method of Levenberg-Marquardt, described in
\cite{NumericalRecipe}, which is based on a steepest descent
method. This first step is done by assuming a surface gravity of $\log
g = 8.0$. This yields an estimate of the effective temperature,
$T_{\rm eff}$, and the solid angle, $\pi \left( R/D \right)^2$ --- or
the radius $R$ of the star since $D$ is known from the trigonometric
parallax. Evolutionary models are then used to obtain the stellar mass
$M$, and a new estimate of the surface gravity, which will be
different from our initial assumption of $\log g=8.0$. The entire
fitting procedure is then repeated until all parameters are
consistent. The uncertainties associated with the fitted parameters
are obtained directly from the covariance matrix of the
Levenberg-Marquardt minimization procedure (see
\citealt{NumericalRecipe}). Here we rely on C/O-core envelope
models\footnote{See
  http://www.astro.umontreal.ca/$\sim$bergeron/CoolingModels.}
similar to those described in \citet{Fontaine2001} with thick hydrogen
layers of $q({\rm H})\equiv M_{\rm H}/M_{\star}=10^{-4}$ for DA stars,
and much thinner hydrogen layers of only $q({\rm H})=10^{-10}$ for DB
stars, given that such thin layers are representative of
helium-atmosphere white dwarfs. Finally, depending on the spectral
type, we use either pure hydrogen, pure helium, or mixed H/He model
atmospheres (see section \ref{sect:effetH}). Examples of the
photometric technique for both a DA and a DBA white dwarf are
illustrated in Figure \ref{fig:techphot}.

\begin{figure}
\includegraphics[width=\columnwidth]{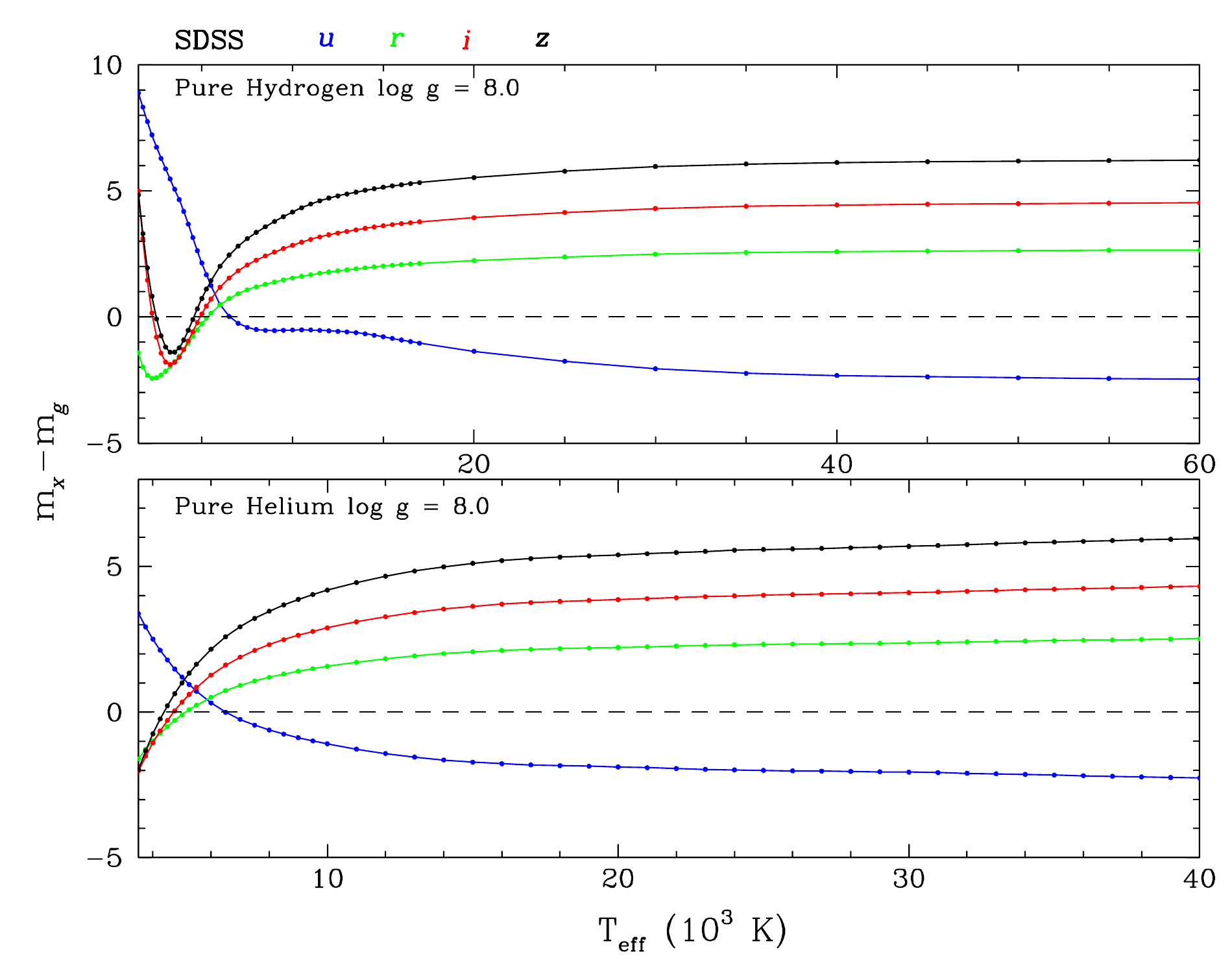}
\caption{Magnitude differences between the SDSS $u$ (blue), $r$
  (green), $i$ (red), and $z$ (black) bands and the SDSS $g$ band, as
  a function of effective temperature, for pure hydrogen (top) and
  pure helium (bottom) models at $\log g=8.0$ (note the difference in
  temperature scales). The dots correspond to individual model
  values.}
\label{fig:lim_phot}
\end{figure}

The precision of the photometric technique is limited by the
sensitivity of the photometric measurements (here SDSS $ugriz$)
to variations in effective temperature. At high $T_{\rm eff}$ values,
the energy distribution sampled by the $ugriz$ photometry is in the
Rayleigh-Jeans regime. This is illustrated in Figure
\ref{fig:lim_phot} where we show magnitude differences (i.e., color
indices) with respect to the $g$ magnitude for all SDSS bandpasses, as
a function of effective temperature and atmospheric composition. In
the case of pure hydrogen atmospheres, the energy
distribution varies considerably up to $T_{\rm eff} \sim 35,000~{\rm
  K}$, and very little above this temperature. This means that the
photometric technique becomes less reliable above 35,000 K for DA
white dwarfs. The situation is similar in the case of DB stars.  For
DA stars, the Balmer jump affects the $u-g$ color index significantly,
which can be observed here as the flat plateau between $\sim$8000 K and
15,000 K.  Also of interest is the behavior below $\sim$5000 K where
collision-induced absorption by molecular hydrogen becomes
important. These two features are of course not observed in pure
helium atmospheres.

\subsection{Effect of the Presence of Hydrogen on the Photometric Solutions}\label{sect:effetH}

\begin{figure}
\includegraphics[width=\linewidth]{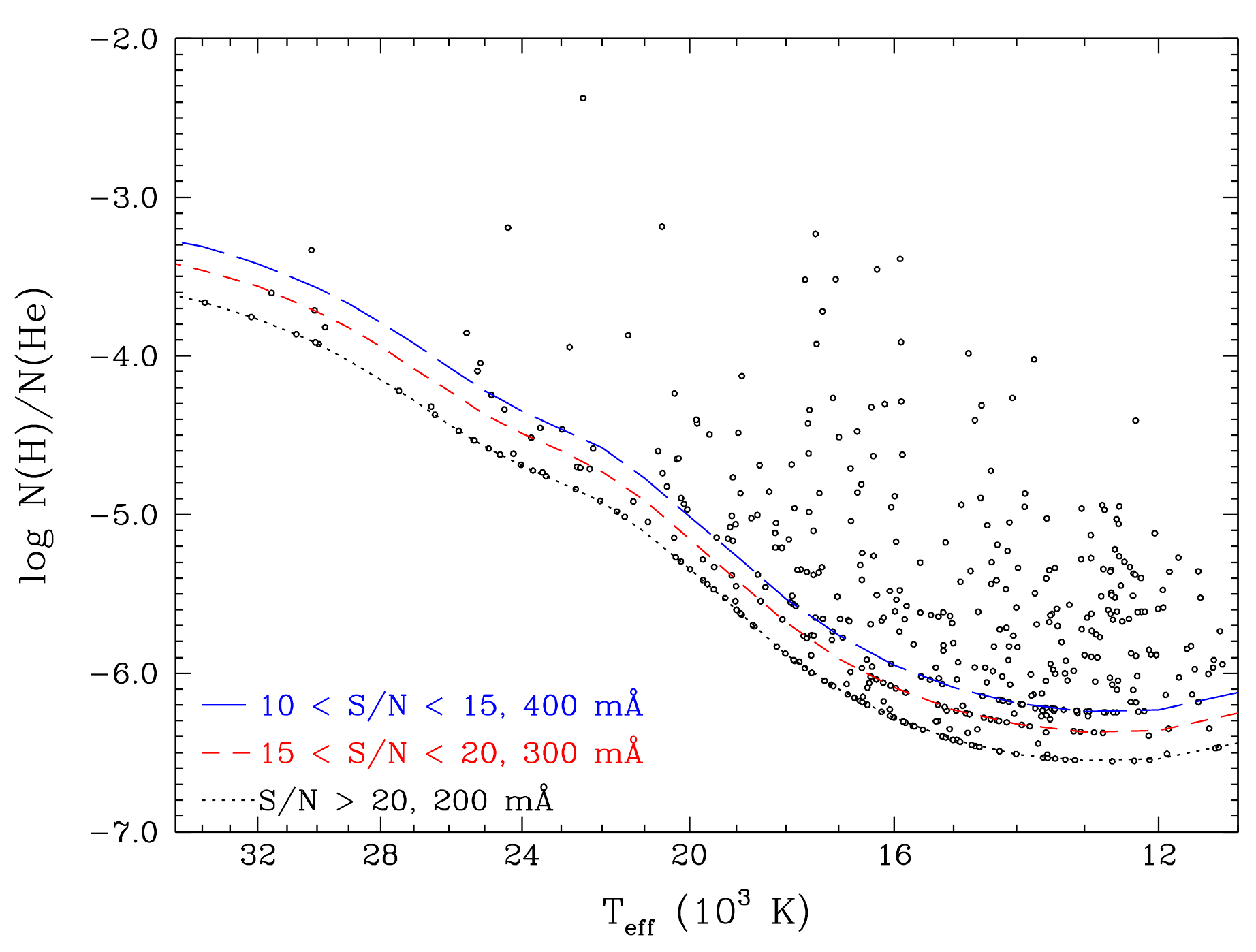}
\caption{Spectroscopic hydrogen abundance as a function of effective
  temperature for the DB and DBA stars in our sample. Also shown are
  the H$\alpha$ detection limits for ${\rm S/N}>20$, $15<{\rm
    S/N}<20$, and $10<{\rm S/N}<15$, corresponding to equivalent
  widths (FWHM) of 200 m\AA~(black, dotted line), 300 m\AA~(red,
  short-dashed line), and 400 m\AA~(blue, long-dashed line),
  respectively.}
\label{fig:Habundance}
\end{figure}

When using the photometric technique, one usually assumes either pure
hydrogen or pure helium atmospheres to determine the stellar
parameters. However, most DB stars contain a certain amount of
hydrogen \citep{Koester2015,Rolland2018}. We show in Figure
\ref{fig:Habundance} the hydrogen abundance (or upper limits) as a
function of effective temperature for all the DB and DBA stars in our
sample. In some cases the hydrogen abundance can be as large as $\log
N(\rm{H})/N(\rm{He})\sim-3$. Since the presence of additional free
electrons in helium-rich atmospheres may affect significantly the
photometric solutions --- see, e.g., Figure 8 of \citet{Dufour05} in
the context of DQ white dwarfs ---, we explore here the effect of the
hydrogen abundance on the atmospheric parameters obtained for DBA
stars using the photometric technique.

\begin{figure*}
\includegraphics[width=\linewidth]{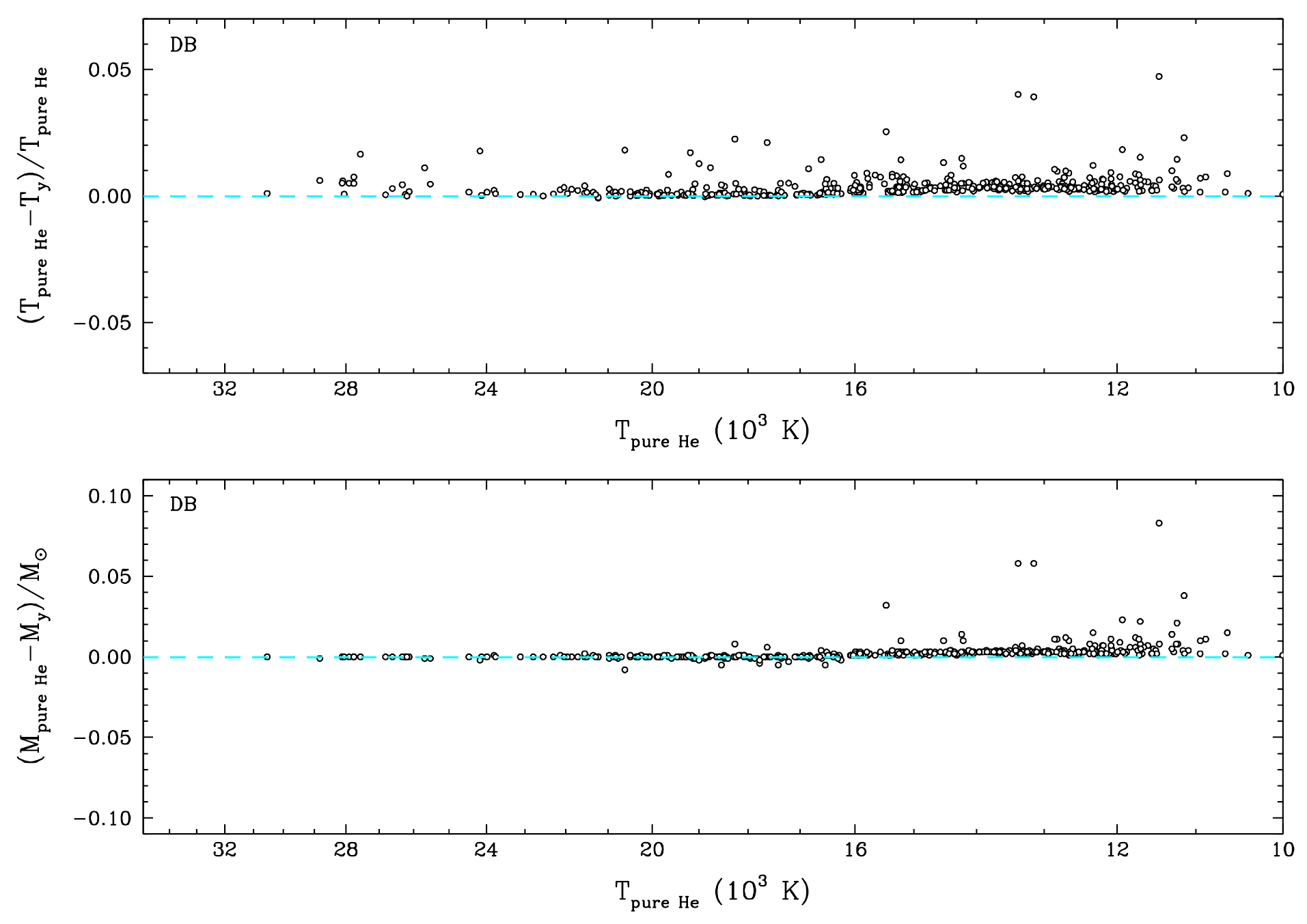}
\caption{Top panel: Effect of the hydrogen abundance on the effective
  temperature determined photometrically, as a function of $T_{\rm
    eff}$. $T_{\rm pure~He}$ is the effective temperature obtained
  from pure helium models, while $T_{y}$ is that obtained by forcing
  the hydrogen abundance to the spectroscopic value. The dashed line
  corresponds to $T_{\rm pure~He}=T_{y}$. Bottom panel: Same as top
  panel, but for stellar masses. The cyan dashed line corresponds to
  $M_{\rm pure~He}=M_y$.}
\label{fig:Hsensitivity}
\end{figure*}

To do so, we use the same procedure as before, but instead of using
pure helium models, we force the hydrogen abundance to the
spectroscopic value. The effects on the effective temperature and
stellar mass are displayed in Figure \ref{fig:Hsensitivity}. The top
panel shows that by using pure helium models, we tend to overestimate
the effective temperature, but not significantly. For most objects,
the difference in temperature is less than 1\% to 2\%, which is
smaller than the uncertainty associated with the photometric
technique. The objects for which the difference in temperature is the
largest, between $\sim$2.5 and 5\%, are all found below $T_{\rm
  eff}\sim14,000~{\rm K}$, and these correspond to the few DBA stars
in this temperature range with the largest hydrogen abundances around
$\log N({\rm H})/N({\rm He}) \sim -4.0$.

The bottom panel of Figure \ref{fig:Hsensitivity} shows the effect on
the mass determinations. Above $T_{\rm eff} \sim 16,000~{\rm K}$, the
effect is completely negligible. Below this temperature, the masses
obtained under the assumption of a pure helium atmosphere are slightly
overestimated, by about 0.01 $M_\odot$ for the bulk of our sample, but
these differences can be as large as 0.05 to 0.1 $M_\odot$ in some
cases. These correspond also to the objects that show the largest
temperature differences in the upper panel. Since the luminosity
$L\propto R^2T_{\rm eff}^4$, a lower temperature implies a larger
radius, or a smaller mass \citep[see also Figure 8 of][]{Dufour05}.

Overall, we conclude that the use of mixed composition atmospheres for
measuring the photometric mass and effective temperature of DBA white
dwarfs yields values very similar to those obtained under the
assumption of pure helium compositions. Nevertheless, to be fully
consistent, we adopt in what follows the spectroscopic hydrogen
abundance to measure the stellar parameters using the photometric
technique.

\subsection{Photometric Results}\label{sect:photresults}

\begin{figure*}
\includegraphics[width=\linewidth]{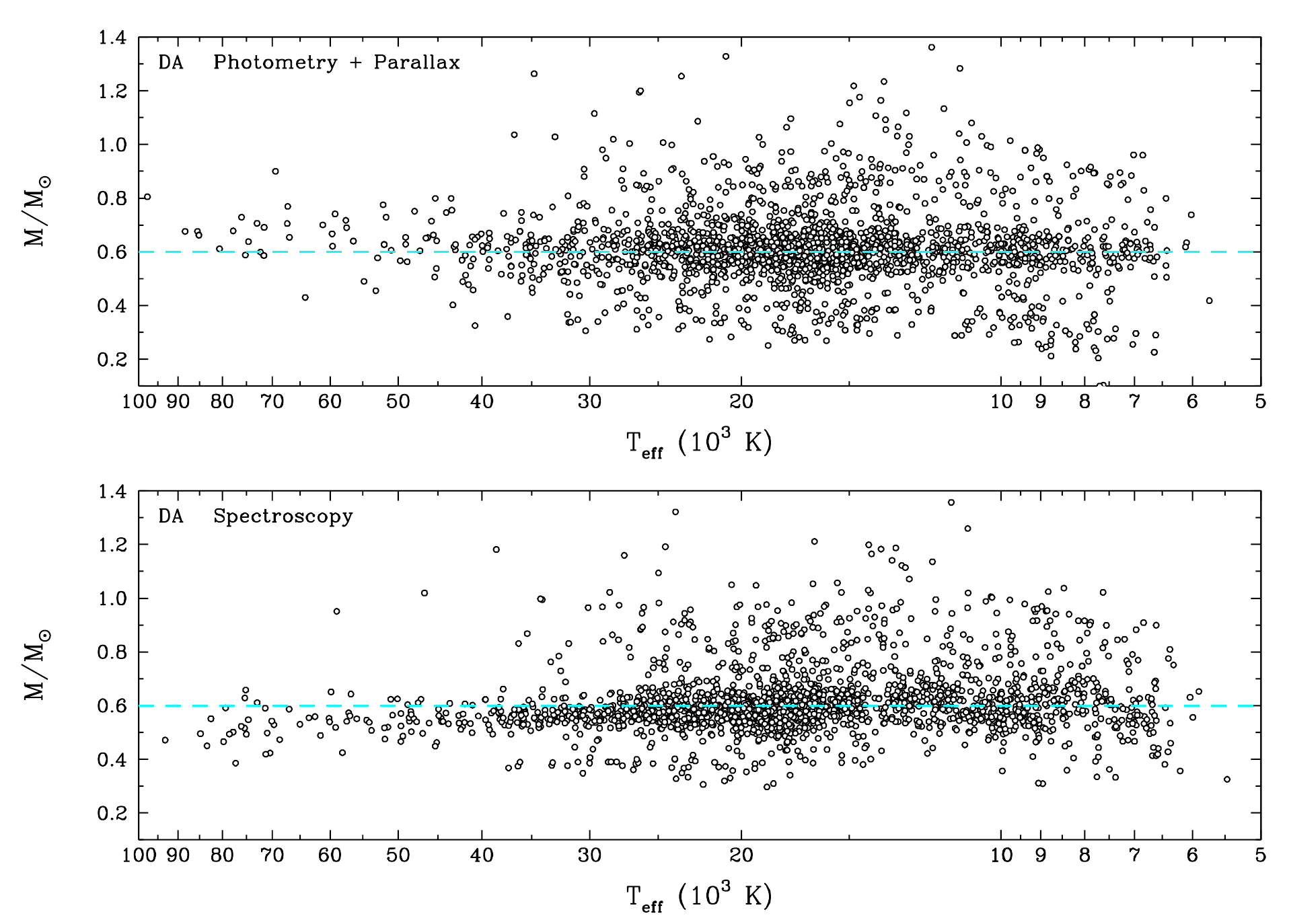}
\caption{Stellar mass as a function of effective temperature obtained
  from photometry (top) and spectroscopy (bottom), for the DA white
  dwarfs in our sample. The dashed cyan line represents a constant
  mass of 0.6 $M_\odot$.}
\label{fig:MvsT_DA}
\end{figure*}

\begin{figure*}
\includegraphics[width=\linewidth]{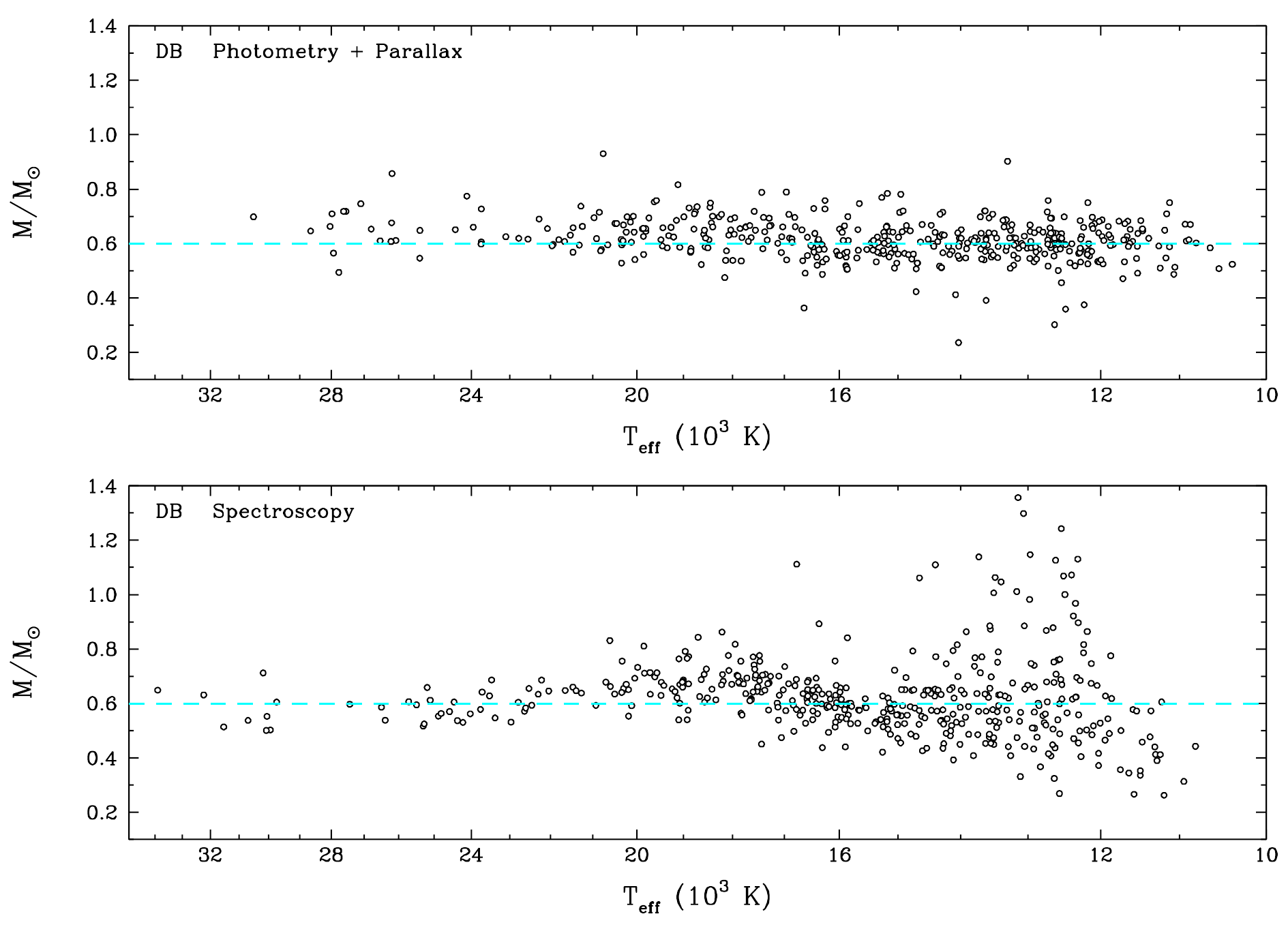}
\caption{Same as Figure \ref{fig:MvsT_DA}, but for the DB white dwarfs in our sample.}
\label{fig:MvsT_DB}
\end{figure*}

Using the photometric technique described in Section
\ref{sect:techphot}, we determined the effective temperature and
stellar mass of every object in our sample. As mentioned in the
previous section, for the DB/DBA stars, the hydrogen abundance (or
limit) was forced to its spectroscopic value. The resulting mass
distributions for the DA and DB stars in our sample are displayed as a
function of effective temperature in the top panels of Figures
\ref{fig:MvsT_DA} and \ref{fig:MvsT_DB}, respectively.

Below $T_{\rm eff}\sim45,000$~K, the mass distribution for the DA
white dwarfs appears well-centered around 0.6 $M_\odot$, regardless of
the effective temperature. This is expected since it is believed that
white dwarfs cool with a constant mass. Above this temperature,
however, the DA stars in our sample appear to have larger than average
masses, around $\sim0.7~M_\odot$. This is probably due to the
limitations of the photometric technique using $ugriz$ data in this
temperature range (see Figure \ref{fig:lim_phot}).  The distribution
of objects in the upper panel of Figure \ref{fig:MvsT_DA} also appears
uniform as a function of $T_{\rm eff}$, i.e., there are no gaps in the
temperature distribution, except perhaps around 12,000 K where there
seems to be a slight depletion of objects. This might be caused by
selection effects in the SDSS, or it could also be due to spectral
evolution mechanisms, such as the transformation of DA into non-DA
white dwarfs resulting from convective mixing in this temperature
range. But such considerations are outside the scope of this paper.

We can also identify a significant number of DA stars with very low
photometric masses ($M\lesssim0.45\ M_\odot$). These objects are most
likely unresolved double degenerate binaries. In these cases, the
object appears overluminous because of the presence of two stars in
the system, and the radius determined photometrically is thus
overestimated. Because of the mass-radius relation, the stellar mass
inferred from this radius is underestimated. These objects will be
discussed further in Sections \ref{sect:Mass} and
\ref{sect:MR}. Finally, the mass distribution also reveals the
existence of high-mass white dwarfs ($M\gtrsim 0.8 M_\odot$). These
high-mass DA stars are usually thought to be the end result of stellar
mergers \citep{Iben1990,Kilic2018}, or alternatively, they can also be
explained as a result of the initial-to-final mass relation
\citep{ElBadry2018}.

As for the DA white dwarfs, the mass distribution for the DB stars
(top panel of Figure \ref{fig:MvsT_DB}) shows a rather uniform
distribution as a function of $T_{\rm eff}$, with no obvious gaps.
Unlike for the DA mass distribution, however, which was centered
around $\sim$0.6 $M_\odot$ regardless of $T_{\rm eff}$, the mean mass
for DB stars appears well-centered around 0.6 $M_\odot$ for $T_{\rm
  eff}\lesssim 16,000~{\rm K}$, but systematically above this value at
higher temperatures.  Also, the most striking feature in the
photometric mass distribution is that there is no significant increase
in mass below $T_{\rm eff} \sim 16,000~{\rm K}$, in sharp contrast
with the spectroscopic mass distributions reported for instance by
\citet{Bergeron2011} and \citet{Koester2015}.  We come back to these
points further in Sections \ref{sect:ivdw} and \ref{sect:Mass}.

We also note the presence of a few low-mass DB white dwarfs
($M\lesssim0.45~M_\odot$). Again, these are most likely unresolved
double degenerate candidates, which will be discussed in more detail
in Sections \ref{sect:Mass} and \ref{sect:MR}.

\section{Spectroscopic Analysis}\label{sect:Spectroscopy}

\subsection{Spectroscopic Technique}\label{sect:techspec}

The most widely used technique to measure the atmospheric parameters
--- $T_{\rm eff}$, $\log g$, and atmospheric composition --- of white
dwarf stars is the so-called spectroscopic technique, which relies on
normalized spectral line profiles. Unlike the photometric technique,
which is the same regardless of the atmospheric composition, the
spectroscopic technique differs slightly depending on the spectral
type of the object. We describe in turn the fitting procedures used
for both DA and DB stars in our SDSS sample.

\subsubsection{DA White Dwarfs}\label{sect:techspecDA}

\begin{figure}
 \includegraphics[width=\columnwidth]{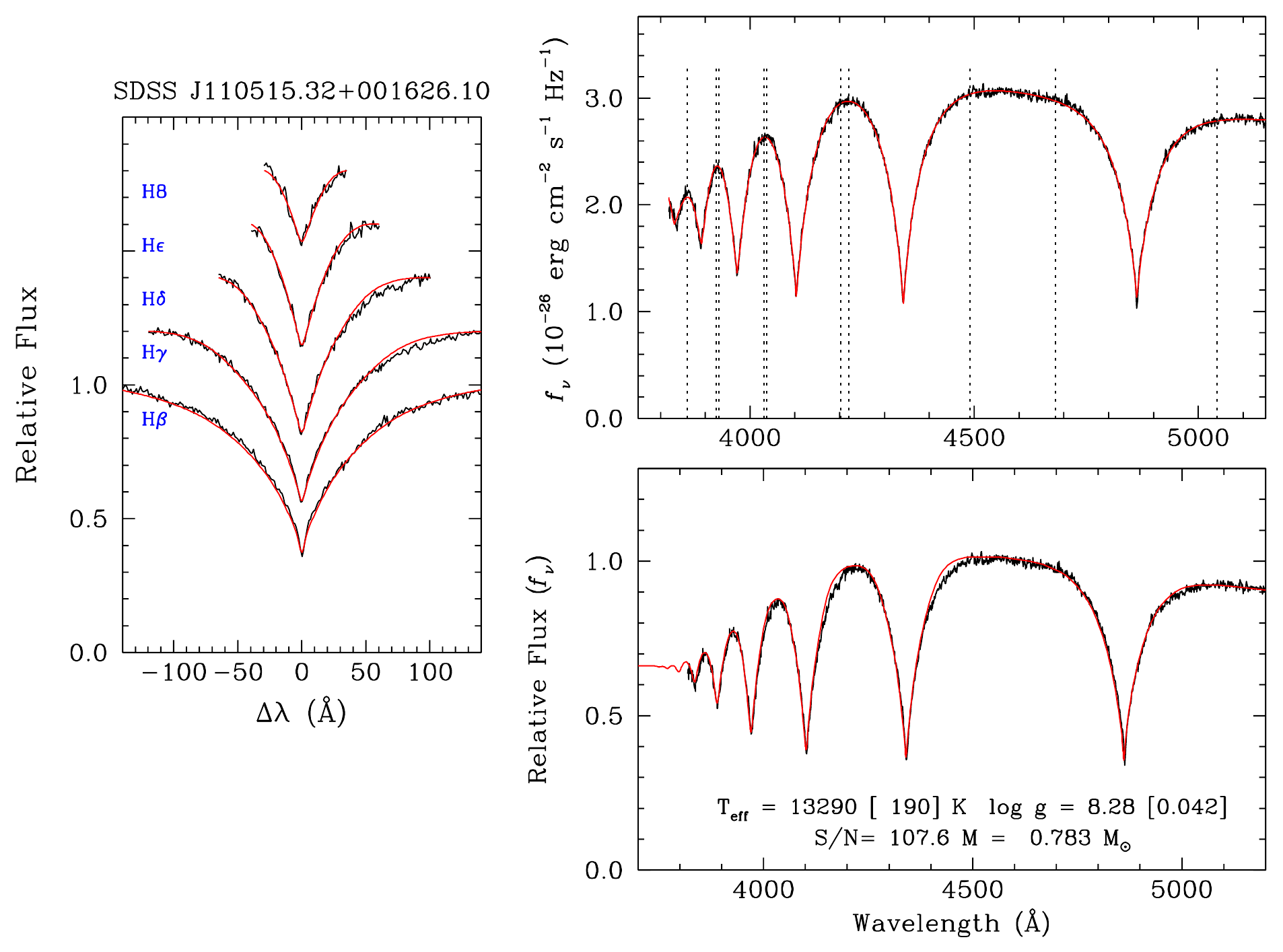}
 \caption{Example of the spectroscopic technique for the DA white
   dwarf SDSS J110515.32+001626.10. Top right panel: The smooth
   fitting function (red) used to define the continuum is plotted over
   the observed spectrum (black). Left panel: Best model fit (red) to
   the observed (normalized) hydrogen line profiles (black).
   Bottom right panel: Final solution (red) superposed on
   the observed spectrum (black), both normalized at 4600 \AA. The
   derived stellar parameters are also given in the figure.}
 \label{fig:techspec_DA}
\end{figure}

For DA white dwarfs, we use a technique similar to that described in
\cite{BSL1992}, \cite{BWL1995}, and \cite{LBH05}. The first step is to
normalize the hydrogen lines, from H$\beta$ to H8, for both the
observed and synthetic spectra, convolved with the appropriate
Gaussian instrumental profile (3 \AA\ FHWM in the case of the SDSS
spectra). The comparison is then carried out in terms of these
normalized line profiles only. In order to properly define the
continuum on each side of the line, we use two different procedures
depending on the temperature range. For $16,000~{\rm K} < T_{\rm eff}
< 9000~{\rm K}$, we fit the entire spectrum using a sum of
pseudo-Gaussian profiles, as they reproduce quite well the spectral
line profiles in this temperature range, an example of which is shown
in the top right panel of Figure \ref{fig:techspec_DA}.  Outside of
this temperature range, we rely on our synthetic spectra to reproduce
the observed spectrum, including a wavelength shift, as well as
several order terms in $\lambda$ (up to $\lambda^6$), to obtain a
smooth fitting function. This is achieved using the
Levenberg-Marquardt method described above.  Since the hydrogen lines
reach their maximum strength around $T_{\rm eff}=14,000~{\rm K}$, both
the cool and hot solutions are tested and the one with the lowest
$\chi^2$ is kept. The resulting best fit is then used to normalize the
line profiles to a continuum set to unity, although the atmospheric
parameters obtained at this point are meaningless because of the high
number of fitting parameters used in the normalization
procedure. However, these $T_{\rm eff}$ and $\log g$ estimates can be
used as a starting point for the full $\chi^2$ minimization procedure
since they are usually quite close to the physical solution. This also
helps us to determine on which side of the maximum line strength our
object is located. In principle, the photometric temperature could
also be used to distinguish between the cool and hot solutions, but we
want our spectroscopic fitting procedure to be as independent as
possible from the photometric approach.

Once the lines are properly normalized, the effective temperature and
surface gravity are determined using the Levenberg-Marquardt
procedure. Finally, the 3D corrections from \cite{Tremblay2013}
are applied to both $T_{\rm eff}$ and $\log g$.  A full example of the
fitting procedure for DA white dwarfs is presented in Figure
\ref{fig:techspec_DA}.

The uncertainties associated with spectroscopic $T_{\rm eff}$ and
$\log g$ values were estimated by \cite{LBH05} for the DA stars in the
Palomar-Green survey. Multiple measurements of the same stars were
used to determine that the overall errors are 1.4\% in $T_{\rm eff}$
and 0.042 dex in $\log g$. Note that these values were obtained using
a spectroscopic sample with ${\rm S/N}>50$. In our SDSS sample,
however, most of our DA spectra have ${\rm S/N} \lesssim 50$ (see
Figure \ref{fig:SN}), with very few objects at higher
values. Nevertheless, we will assume here the same uncertainties as
those of Liebert et al., but we keep in mind that these are most
likely underestimated.

\subsubsection{DB/DBA White Dwarfs}\label{sect:techspecDB}

The spectroscopic technique used for the DB/DBA white dwarfs differs
slightly from that used for DA stars since there is a third parameter
to measure: the hydrogen abundance. We use a technique similar to that
described in \cite{Bergeron2011}. The normalization procedure for the
DB spectra relies on our synthetic spectra to obtain a smooth fitting
function used to determine the continuum, as described above for the DA
stars. Again, we find a solution on each side of the maximum line
strength, which occurs near 25,000 K for DB white dwarfs, and the
solution with the smallest $\chi^2$ is used to define the
continuum. As before, this normalization procedure uses too many
fitting parameters for the values of $T_{\rm eff}$, $\log g$, and
$\log N({\rm H})/N({\rm He})$ to be meaningful.

\begin{figure}
\includegraphics[width=\columnwidth]{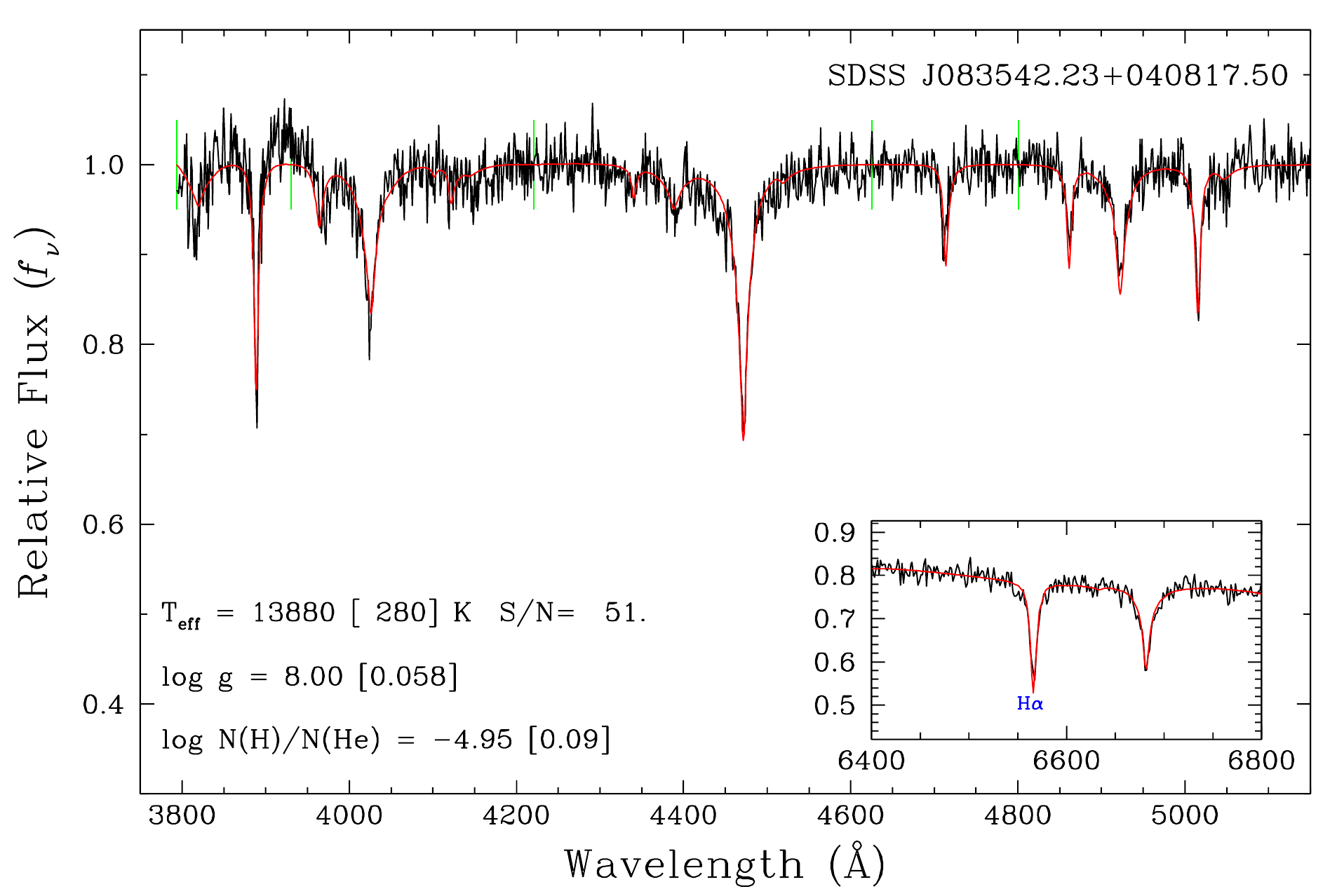}
\caption{Example of the spectroscopic technique for the DBA white
  dwarf SDSS J083542.23+040817.50. The best fit (red) is plotted over
  the normalized observed spectrum (black). The inset shows the region
  near H$\alpha$ used to determine the hydrogen abundance, or upper
  limits.  The derived atmospheric parameters are also given in
  the figure.}
\label{fig:techspec_DB}
\end{figure}

Once the observed spectrum is normalized, the first step is to obtain
an estimate of the effective temperature and surface gravity, using
the blue part of the spectrum ($\lambda=3750-5150$ \AA). Keeping those
parameters fixed, the hydrogen abundance is obtained by fitting the
region near H$\alpha$ ($\lambda=6400-6800$ \AA). For some stars, this
part of the spectrum is problematic, so the hydrogen abundance is
determined using H$\beta$ instead. The entire procedure is then
repeated in an iterative fashion, until the value of $N({\rm
  H})/N({\rm He})$ has converged. In several cases, only upper limits
on the hydrogen abundance could be obtained based on the absence of
H$\alpha$ within the detection limit (see Figure
\ref{fig:Habundance}). An example of the fitting procedure for a
typical DBA white dwarf is shown in Figure \ref{fig:techspec_DB}.

\cite{Bergeron2011} determined, in the same manner as \cite{LBH05},
that the overall error on the effective temperature and surface
gravity were 2.3\% and 0.052 dex, respectively. Although we will be
using those estimates, it should be noted that they determined those
values with much higher signal-to-noise spectra than what is use in
the present study (${\rm S/N} \gtrsim 50$ in \citealt{Bergeron2011} vs
${\rm S/N} > 10$ here, see Figure \ref{fig:SN}). Therefore, our
uncertainties are most likely underestimated.

\cite{Cukanovaite2018} calculated a series of 3D hydrodynamical white
dwarf atmosphere models with pure helium compositions, similar to
those for DA stars by \cite{Tremblay2013}. They also published 3D
corrections to be applied to 1D spectroscopic solutions, which are
found to be important in the range $T_{\rm eff}\sim 17,000~{\rm K} -
20,000~{\rm K}$. However, since it is expected that the presence of
hydrogen will affect these corrections, and that such calculations are
currently underway, we refrain from applying any correction to our
spectroscopic solutions at this stage.  We will keep this in mind,
however, when we compare the photometric and spectroscopic results in
Section \ref{sect:comparison}. Note that these more realistic 3D
hydrodynamical models should not affect in any way the photometric
analyses presented above for both DA and DB white dwarfs.

\subsection{van der Waals Broadening in DB White Dwarfs}\label{sect:ivdw}

A well-known problem in the case of DB white dwarfs is the apparent
increase in $\log g$, or mass, at low effective temperatures ($T_{\rm
  eff} \lesssim 15,000~{\rm K}$). This phenomenon has been reported
repeatedly, for instance, in \cite{Beauchamp1996},
\cite{Bergeron2011}, and \cite{Koester2015}.  The photometric mass
distribution displayed in the top panel of Figure \ref{fig:MvsT_DB}
reveals that this increase in mass occurs only when the atmospheric
parameters are determined using the spectroscopic technique, a
conclusion also reached by \citet{Tremblay2019}. A similar phenomenon
was also observed in the $\log g$ distribution of cool DA stars ---
the so-called high-$\log g$ problem --- but in this case,
\citet{Tremblay2013} showed that the problem lies in the use of the
mixing-length theory to treat convective energy transport, and that
more realistic 3D hydrodynamical calculations could solve this
high-$\log g$ problem.

\begin{figure*}
\includegraphics[width=\linewidth]{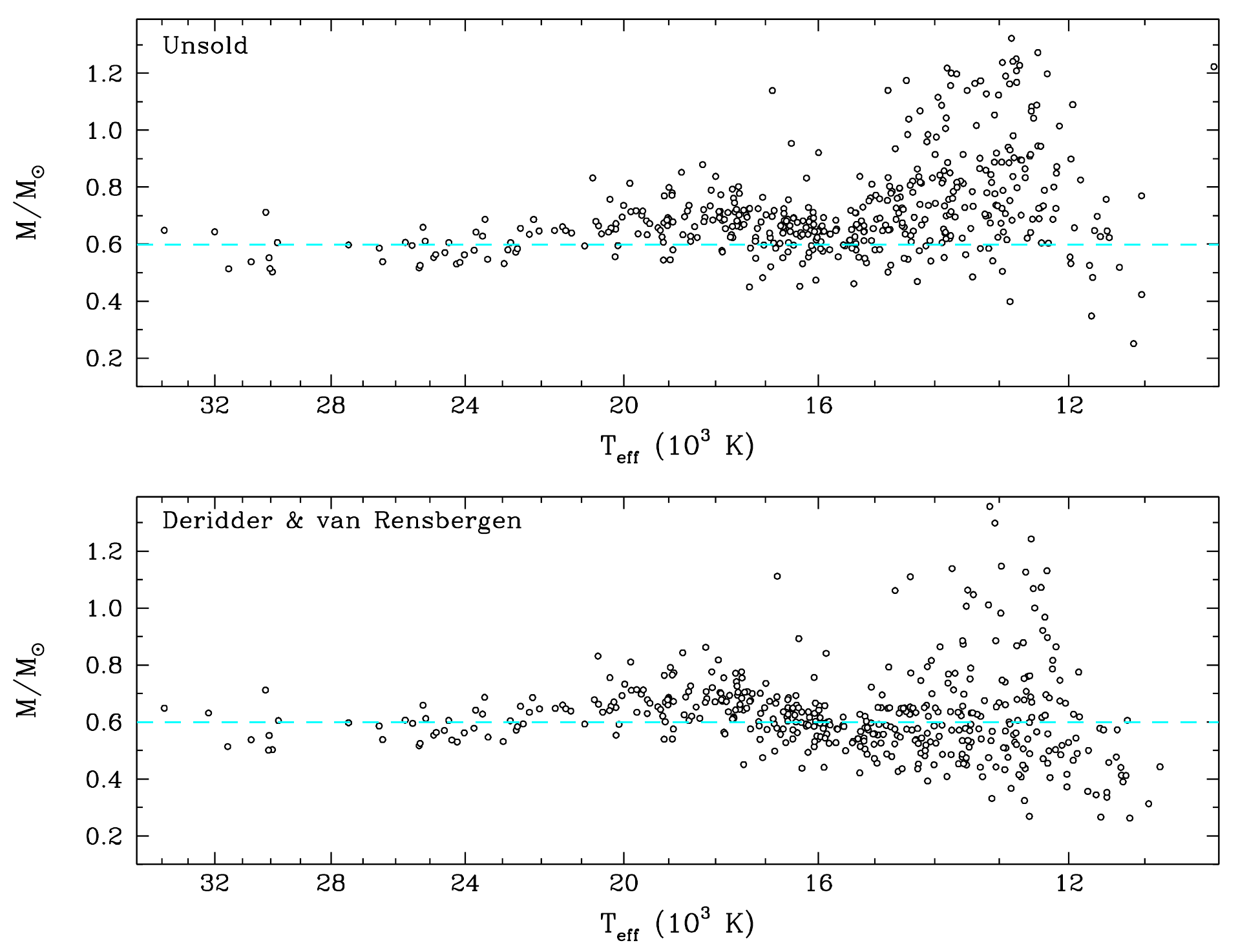}
\caption{Spectroscopic mass distribution as a function of effective
  temperature for the DB white dwarfs in the SDSS sample, using the van
  der Waals broadening theory of \citet[][top]{Unsold1955} and Deridder
  \& van Rensbergen (bottom; as defined in Section
  \ref{sect:modelDB}). The dashed line in both panels corresponds to a
  constant mass of 0.6 $M_\odot$.}
\label{fig:ivdw}
\end{figure*}

The high-$\log g$ values inferred for cool DB white dwarfs most likely
have a different origin, however, since in the temperature regime where
the problem is observed, convection is almost completely adiabatic
\citep{Cukanovaite2018}. Instead, it has been generally argued that
van der Waals broadening was the source of the problem
(\citealt{Bergeron2011} and references therein). In their paper,
\citet[][see also \citealt{Rolland2018}]{Bergeron2011} erroneously
mention that they used the Deridder \& van Rensbergen theory to treat van
der Waals broadening, as defined in Section \ref{sect:modelDB}, while
they were in fact relying on the more simple theory of
\cite{Unsold1955}\footnote{In this case, the Unsold theory was used
  for every line, but the total width was still $\omega_{\rm
    neutral}={\rm max}(\omega_{\rm resonance},\omega_{\rm vdW})$.}.
To clarify this situation, we fitted all the DB stars in our SDSS sample
using both the Unsold and the Deridder \& van Rensbergen theories, the
results of which are displayed in Figure \ref{fig:ivdw}.

Our results using the Unsold theory can be compared directly with
those shown, for instance, in Figure 6 of
\citet{Rolland2018}. Although the sample analyzed by Rolland et al.~is
significantly smaller than our SDSS sample, both distributions are
qualitatively similar. In particular, they both show a large increase
in mass below $T_{\rm eff}\sim15,000$~K. By using instead the Deridder
\& van Rensbergen theory (bottom panel of Figure \ref{fig:ivdw}), we
still find white dwarfs with large masses ($M \sim 1.2-1.3~M_\odot$)
at low temperatures, but more importantly, the mass of the bulk of our
sample has been reduced from a value above 0.6 $M_\odot$ to a value
significantly below this mark. Furthermore, the mass distribution now
contains several objects with low masses ($M \lesssim
0.45\ M_\odot$). Finally, the scatter at low temperatures remains
large with both theories. Note that the stellar masses are relatively
unaffected above $\sim$16,000 K, where Stark broadening dominates.

The results displayed in Figure \ref{fig:ivdw} indicate that the
theory of van der Waals broadening still requires significant
improvement before any meaningful spectroscopic analysis of cool
DB/DBA white dwarfs can be achieved. For instance, our particular
choice (see Section \ref{sect:modelDB}) --- based on the results of
\citet{Lewis1967} --- to use $\omega_{\rm neutral}={\rm
  max}(\omega_{\rm resonance},\omega_{\rm vdW})$ instead of the sum of
the two contributions, might not be appropriate. \citet{Mullamphy1991}
indeed found that a simple sum, $\omega_{\rm neutral}=\omega_{\rm
  resonance}+\omega_{\rm vdW}$, provided a better agreement between
theory and experiment. However, the inadequacy in the treatment of
He~\textsc{i} line broadening by neutral particles is outside the
scope of this paper, and further improvements will be explored
elsewhere. For the lack of a better theory for van der Waals
broadening, we use in the remainder of our analysis the Deridder \&
van Rensbergen theory, as defined in Section \ref{sect:modelDB}.

\begin{figure*}
\includegraphics[width=1.0\linewidth]{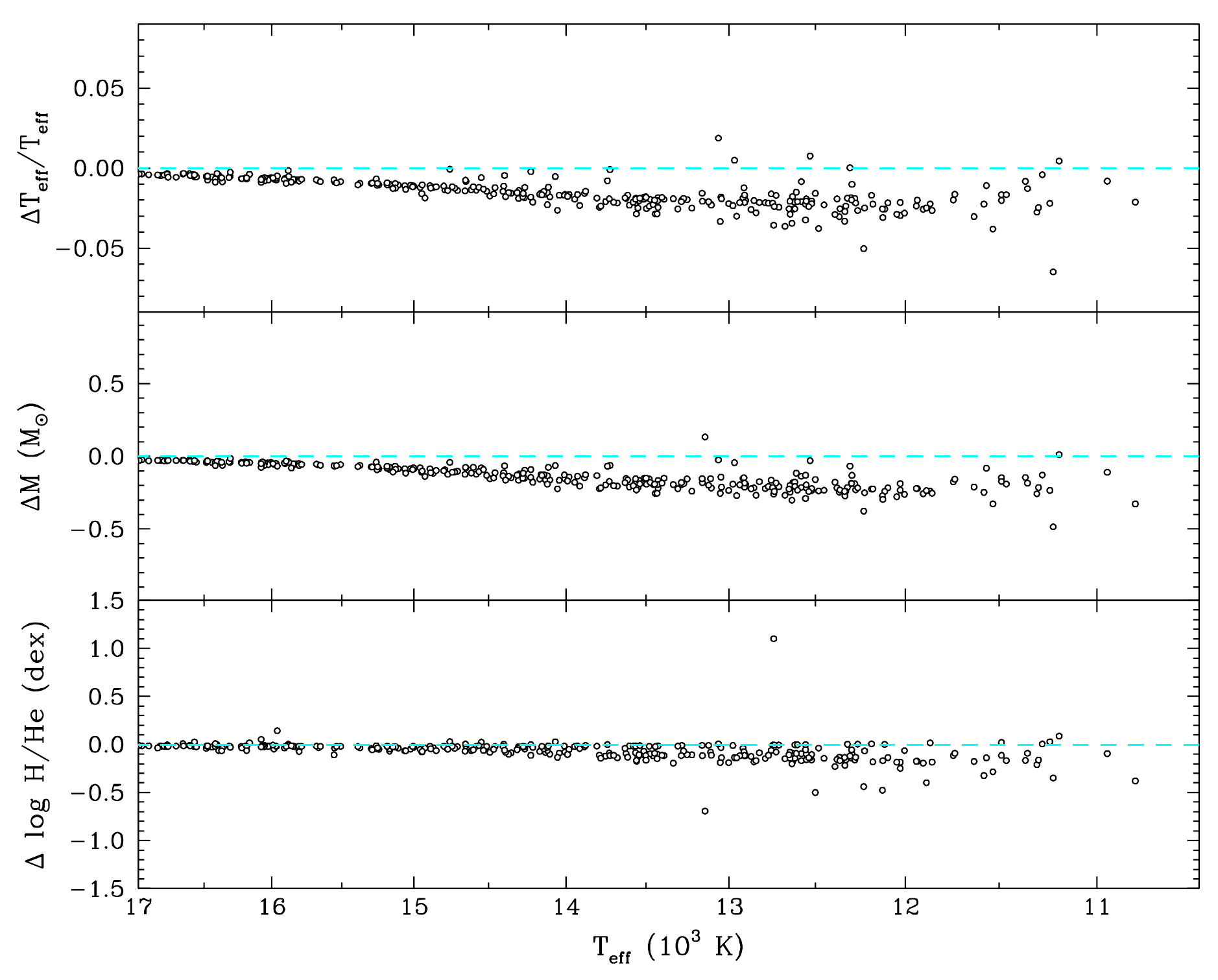}
\caption{Differences in effective temperature, mass, and hydrogen
  abundances as a function of $T_{\rm eff}$ obtained between models using
  the van der Waals broadening theory of Deridder \& van Rensbergen (as
  defined in Section \ref{sect:modelDB}) and the \citet{Unsold1955}
  theory.}
\label{fig:ivdw_effect}
\end{figure*}

For completeness, we show in Figure \ref{fig:ivdw_effect} the
differences in effective temperature, mass, and hydrogen abundance
obtained from models using the Deridder \& van Rensbergen and Unsold
theories, but only for the objects in our sample with $T_{\rm eff} <
17,000~{\rm K}$. The maximum differences occur near $T_{\rm eff} \sim
13,000~{\rm K}$, and are of the order of 2.5\% in temperature, 0.2
$M_\odot$ in mass, but only 0.2 dex in $\log N({\rm H})/N({\rm He})$.

\subsection{Spectroscopic Results}\label{sect:specresults}

Using the spectroscopic techniques described in section
\ref{sect:techspec}, we measured the atmospheric parameters of all the
DA and DB stars in our SDSS sample. Stellar masses were then obtained
by converting the spectroscopic $\log g$ values into mass using the
same evolutionary models as those described in Section
\ref{sect:techphot}. We discuss these results in turn.

The spectroscopic mass distribution for the DA white dwarfs is shown
as a function of effective temperature in the bottom panel of Figure
\ref{fig:MvsT_DA}, which can be contrasted with the photometric
distribution in the upper panel. Both distributions reveal several
high-mass white dwarfs ($M \gtrsim 0.8~M_\odot$), which, as mentioned
above, are believed to be the product of stellar mergers, or
alternatively, the result of the initial-to-final mass relation. We
also note the presence of low-mass white dwarfs ($M\lesssim
0.4~M_\odot$) in both distributions, most likely unresolved double
degenerate binaries (see Section \ref{sect:Mass}).

The spectroscopic mass distribution of DA stars shows a rather uniform
distribution as a function of $T_{\rm eff}$, with the exception of an
obvious gap near 14,000 K. Similar but less obvious gaps were
mentioned in the photometric mass distribution as well (see Section
\ref{sect:photresults}), but these occur at different temperatures and
are probably unrelated. The gap observed in the spectroscopic mass
distribution of DA stars actually corresponds to the temperature where
hydrogen lines reach their maximum strength. If our models predict
lines that are stronger than what is actually observed, the
spectroscopic technique will push the stars away from the maximum,
either on the cool or hot side, to match the observed spectrum, as
observed here. On the other hand, if the predicted lines are weaker
than observed, stars would tend to accumulate near $T_{\rm eff} \sim
14,000~{\rm K}$. Both situations are illustrated in Figure 3 of
\cite{BWL1995} where the mass distribution of bright DA stars is shown
as a function of effective temperatures for different
parameterizations of the MLT, which affect significantly the predicted
maximum line strength. Also of importance are the adopted Stark
broadening profiles used in the synthetic spectrum calculations ---
those of \citet{Tremblay2009} in our models. The gap observed in
Figure \ref{fig:MvsT_DA} suggests that Stark broadening or the
treatment of convection, or even both, might need to be revisited.

There is even a third alternative explanation for the presence of this
gap. Indeed, \citet{Genest2014} reported a similar deficit of objects
near 14,000~K in the spectroscopic temperature distribution of DA
stars from the SDSS (see their Section 3.4 and their Figures 14 and 15),
while an {\it accumulation} of objects was observed instead when using
the DA spectra from \citet{Gianninas2011}. Furthermore, they also
found that while the mass distribution as a function of $T_{\rm eff}$
followed a constant mean value of $\sim$0.6 $M_\odot$ when using the
Gianninas sample, the spectroscopic masses at higher temperatures were
lower than this canonical value when using SDSS spectra, as also
observed here in the lower panel of Figure \ref{fig:MvsT_DA} (for
$T_{\rm eff} \gtrsim 30,000~{\rm K}$). Based on these results,
Genest-Beaulieu \& Bergeron concluded that the SDSS spectra may still
suffer from calibration issues.

The spectroscopic mass distribution for the DB stars, presented in the
bottom panel of Figure \ref{fig:MvsT_DB}, is much more complex
than in the case of DA white dwarfs. The most striking detail about
this distribution is the very large scatter in mass for $T_{\rm eff} \lesssim
15,000~{\rm K}$. This was also discussed in Section \ref{sect:ivdw},
and the most likely cause of this scatter is the improper treatment of
van der Waals broadening in our model atmospheres. Also, contributing
to this scatter is the lack of sensitivity of the spectroscopic
technique below $T_{\rm eff} \sim 12,000~{\rm K}$, when the helium
lines become too weak (see, e.g., Figure 3 of \citealt{Rolland2018}).

The spectroscopic mass distribution for DB stars also shows a
significant increase in mass of $\sim$0.1~$M_\odot$ in the range
$20,000~{\rm K} > T_{\rm eff} > 16,000~{\rm K}$. As discussed above,
this corresponds to the temperature range where 3D hydrodynamical
effects are expected to become important in pure helium DB stars
\citep{Cukanovaite2018}.  Cukanovaite et al.~showed that the largest
differences in spectroscopic $\log g$ values inferred from both 1D and
3D models occur at $T_{\rm eff} \sim 18,000~{\rm K}$, and that 1D
models tend to overestimate the surface gravities, and thus masses, as
observed here. But as discussed above, the photometric masses also lie
above 0.6 $M_\odot$ in the same temperature range. We come back to
this point in the next section.

\section{Comparison of Photometric and Spectroscopic Atmospheric Parameters}\label{sect:comparison}

In this section we compare the effective temperatures and stellar masses
obtained from the photometric and spectroscopic techniques (Sections
\ref{sect:Photometry} and \ref{sect:Spectroscopy}) for both the DA and
DB white dwarfs in our SDSS sample.

\subsection{Effective Temperatures}\label{sect:Teff}

\begin{figure*}
\includegraphics[width=\linewidth]{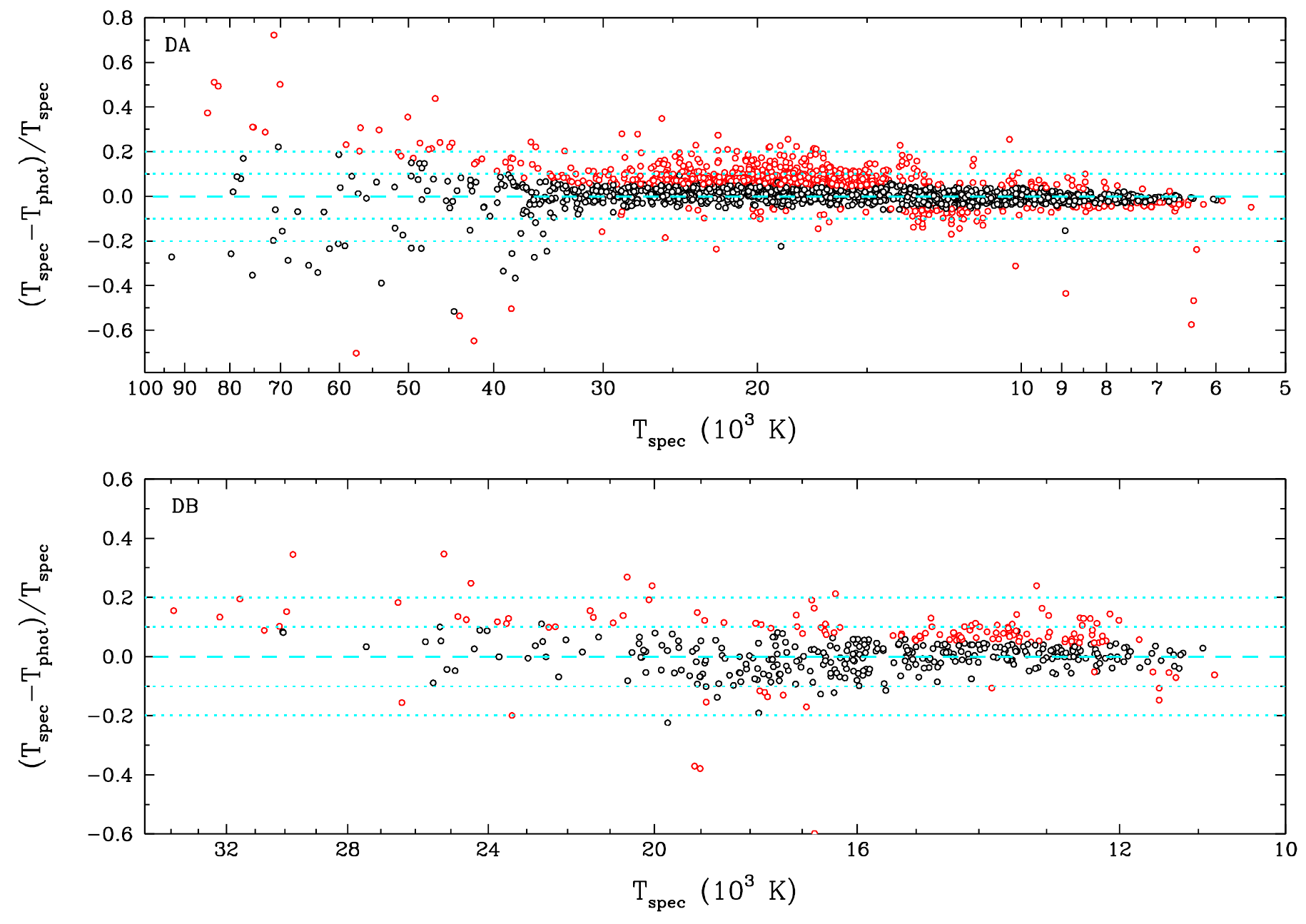}
\caption{Differences between spectroscopic and photometric effective
  temperatures as a function of $T_{\rm spec}$ for the DA (top) and DB
  (bottom) white dwarfs in our SDSS sample. Objects with temperature
  differences within (outside) the 1$\sigma$ confidence level are
  shown in black (red). The dashed line corresponds to $T_{\rm spec} =
  T_{\rm phot}$, while the dotted lines show $\pm 10$\% and $\pm 20$\%
  differences in temperature.}
\label{fig:deltaTeff}
\end{figure*}

The differences between the spectroscopic and photometric effective
temperatures --- $T_{\rm spec}$ and $T_{\rm phot}$ ---
for the DA white dwarfs in our sample are shown as a function of
$T_{\rm spec}$ in the top panel of Figure \ref{fig:deltaTeff}. Here
and below, we use black (red) symbols to indicate white dwarfs whose
temperature estimates are within (outside) the 1$\sigma$
confidence level, where $\sigma$ is defined as the combined
photometric and spectroscopic uncertainties,
$\sigma^2\equiv\sigma^2_{T_{\rm phot}}+\sigma^2_{T_{\rm spec}}$.
Using this definition, we find that 63.3\% of the DA white dwarfs
in our sample have temperature estimates that agree within 1$\sigma$. This
is somewhat lower than expected from Gaussian statistics (68\%), but
as mentioned in Sections \ref{sect:techspecDA} and
\ref{sect:techspecDB}, we are most likely underestimating the
uncertainties associated with our \textit{spectroscopic} parameters,
for both the DA and DB white dwarfs. 

Despite this overall agreement between the photometric and
spectroscopic temperatures, we can observe some obvious systematic
effects in the top panel of Figure \ref{fig:deltaTeff}, which depend
on the range of temperatures considered. For instance, below $T_{\rm
  eff}\sim 14,000$~K, as much as 76\% of the objects agree within
1$\sigma$, with no obvious systematic trend. Above this temperature,
however, only 58\% of the sample is within 1$\sigma$, and more
importantly, spectroscopic temperatures are about 5\% to 10\% higher
than those inferred from photometry. This systematic offset also
appears to be fairly constant through the entire temperature range
above 14,000 K. A similar offset has been reported before by
\citet[][see their Figure 20 and the discussion in their Section
  4]{Genest2014} using similar SDSS photometric and spectroscopic
data. The authors note that this effect is also observed (see their
Figure 22) when using the DA spectra from \citet{Gianninas2011}, and
is thus not related to the particular use of SDSS
spectra. \citet{Tremblay2019} also observed a similar systematic
offset using {\it Gaia} photometry, but in their case, the offset is
seen for all temperatures.

\begin{figure*}
\includegraphics[width=\linewidth]{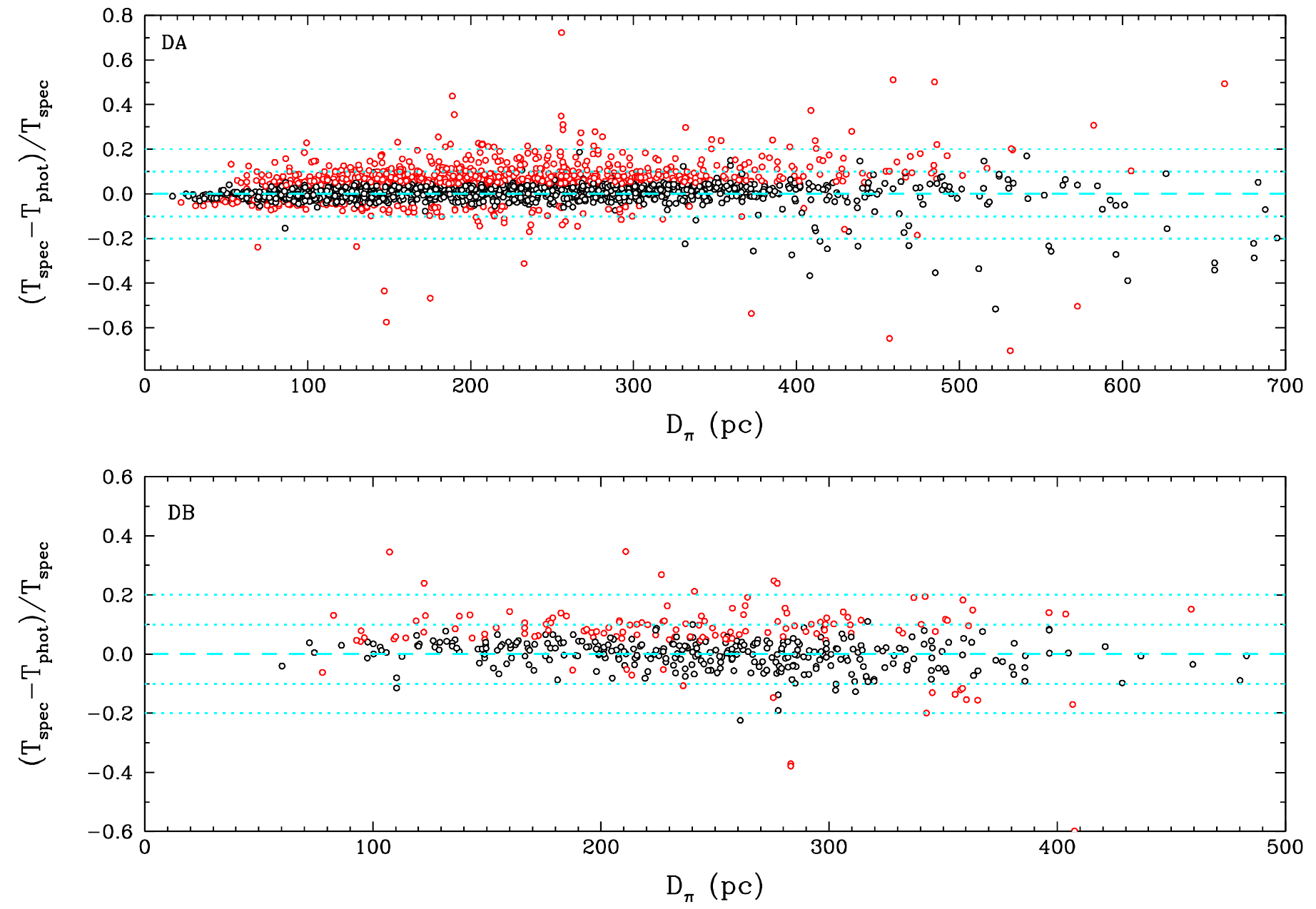}
\caption{Same as Figure \ref{fig:deltaTeff} but as a function of parallactic distances.}
\label{fig:dTeffvsD}
\end{figure*}

\begin{figure*}
\includegraphics[width=\linewidth]{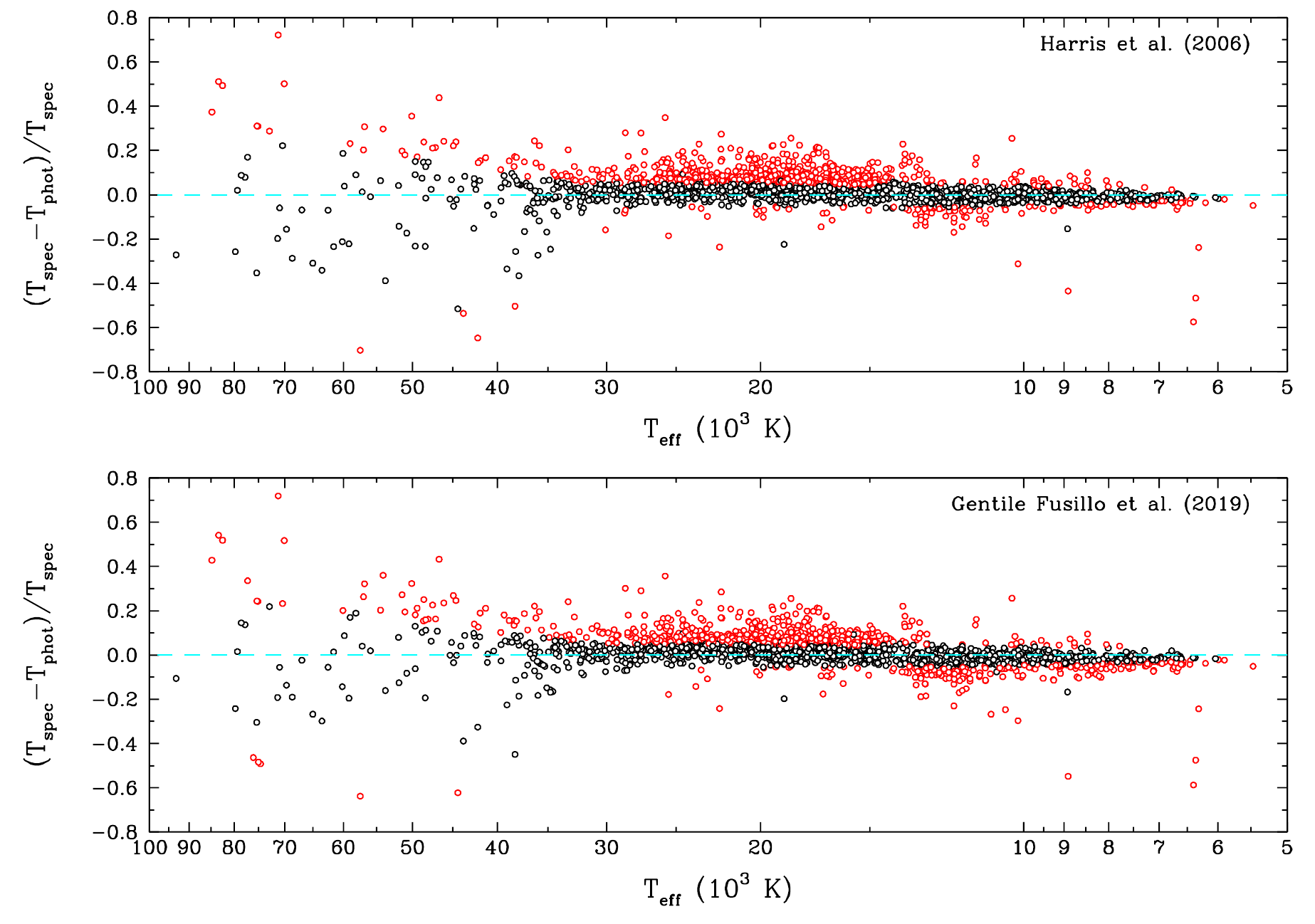}
\caption{Same as the upper panel of Figure \ref{fig:deltaTeff} (DA stars only), where
  interstellar extinction is taken into account by following the
  approach described by \citet[][upper panel]{Harris2006} and by
  \citet[][lower panel]{Gentile2019}.}
\label{fig:reddening}
\end{figure*}

Since interstellar reddening is important for the SDSS sample, in
particular for the hotter and thus intrinsically more luminous and
more distant objects, we explore in Figure \ref{fig:dTeffvsD} the
exact same results as in Figure \ref{fig:deltaTeff}, but this time as
function of the parallactic distance $D_\pi$. We can see here that the
systematic offset in temperature is not a function of distance, and
more importantly, it is also present below 100 pc where interstellar
reddening is negligible according to the dereddening procedure
described in \citet{Harris2006}. \citet{Gentile2019} proposed an
alternative dereddening procedure (see their Section 4) that differs
slightly from that used by Harris et al., in particular for $D<100$
pc. We explore in Figure \ref{fig:reddening} the differences between
these two recipes, but only for the DA stars in our sample. As can be
seen, the results are virtually identical. Actually, the fraction of
white dwarfs whose temperature estimates are within the 1$\sigma$
confidence level decreases from a value of 63.3\% with the Harris et
al.~procedure, to a value of 61.2\% with the Gentile Fusillo et
al.~approach. We thus conclude that our dereddening procedure
described in Section \ref{sect:techphot} is probably reliable, and
that it is not the source of the systematic temperature discrepancy
observed in Figure \ref{fig:deltaTeff}.

One possible reason for the temperature discrepancy might be related
to the physics of line broadening theory, which is likely to affect
more importantly the spectroscopic parameters than those obtained from
photometry, a solution also proposed by \citet{Tremblay2019}. For
instance, \cite{Genest2014} compared in their Figure 23 the
photometric and spectroscopic temperatures for the DA stars in the
Gianninas et al.~sample, but by using model spectra calculated with
the Stark profiles of \cite{Lemke1997} --- with twice the value of the
critical electric microfield $\beta_{\rm crit}$ in the Hummer-Mihalas
occupation probability formalism (see \citealt{BSL1992} and references
therein) --- instead of those of \cite{Tremblay2009} used in our
analysis. They obtained a much better agreement with these older
profiles above $T_{\rm eff} \sim 20,000$~K, although other systematic
effects were introduced between 13,000 K and 19,000 K.  Nevertheless,
these results strongly suggest that Stark broadening probably needs
further improvements, for instance along the lines of the promising
work of \citet{Gomez17}.

The region around $T_{\rm eff} = 14,000~{\rm K}$ in Figure
\ref{fig:deltaTeff} also appears problematic. As mentioned above, this
corresponds to the temperature at which the hydrogen lines reach their
maximum strength. The spectroscopic solutions in this region are
particularly sensitive to the treatment of atmospheric convection, to
3D hydrodynamical effects, to the physics of Stark broadening, and
even to flux calibration issues. All these effects are responsible for
moving the objects around the region where the lines reach their
maximum strength, and for producing the increased scatter near 14,000
K in Figure \ref{fig:deltaTeff}. It is even possible that for some
objects, the spectroscopic technique did not pick the correct
solution, cool or hot, in particular when both are close to the
photometric temperature, in which case it is virtually impossible to
discriminate between both solutions.

Finally, we note in Figure \ref{fig:deltaTeff} that the scatter
increases significantly for $T_{\rm spec}>40,000~{\rm K}$. This is
obviously caused by the lack of sensitivity of the $ugriz$ photometry
in the Rayleigh-Jeans regime (see Figure \ref{fig:lim_phot}). This
scatter could possibly be reduced if we were to extend our set of
photometric data towards shorter wavelengths, by using {\it Galex}
photometry, for instance.

\begin{figure}
\includegraphics[width=\linewidth]{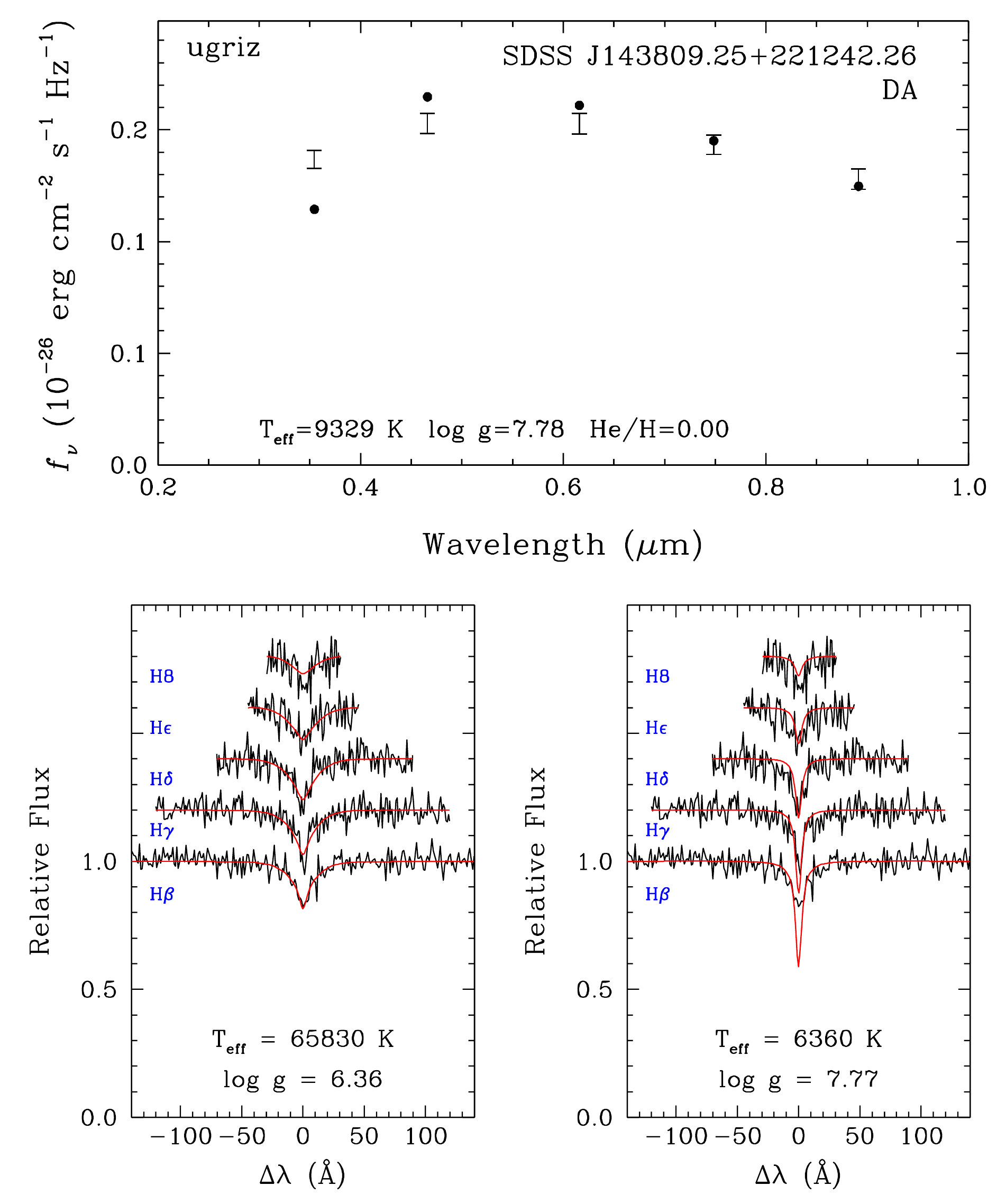}
\caption{Top panel: Best photometric fit to SDSS J143809.25+221242.26
  --- a double degenerate DA+DC candidate --- assuming a single DA star
  with a pure hydrogen atmosphere. Bottom panels: Best spectroscopic
  fits for the same object showing the hot (left) and cool (right)
  solutions.}
\label{fig:DA+DC}
\end{figure}

In some cases, the large differences between the photometric and
spectroscopic solutions may be indicative of the presence of an
unresolved double degenerate binary. An example of such a candidate in
our sample is SDSS J143809.25+221242.26, displayed in Figure
\ref{fig:DA+DC}. The best photometric fit indicates a temperature of
9329 K and a rather low value of $\log g=7.78$. We also note
that this is a particularly bad photometric fit, especially at
$ugr$. If we drop the $u$ bandpass, we can achieve a much better fit
(not shown here) with $T_{\rm eff}=8700$ K and $\log g=7.63$, in good
agreement with the pure hydrogen fit from the Montreal White Dwarf
Database \citep{MWDD}, based on Pan-STARRS $grizy$ photometry ($T_{\rm
  eff} = 8444$ K, $\log g=7.56$). In the bottom panels of Figure
\ref{fig:DA+DC}, we show our best spectroscopic fits for the same
object assuming a hot and a cool solution. The hot solution provides a
much better fit to the observed hydrogen line profiles, while the cool
solution predicts lines that are way too deep; both spectroscopic
temperatures are significantly different from the photometric value,
however.  This large temperature discrepancy as well as the distortion
of the photometric fit suggest that SDSS J143809.25+221242.26 is an
unresolved degenerate binary composed of a DA+DC, where the
hydrogen lines of the DA component are being diluted by the DC white
dwarf in the system. Note that \citet{DR10} reported values of $T_{\rm
  eff} = 10,049$ K and $\log g=9.04$ for the same object. A more
detailed analysis of this system, and other such binary candidates in
our sample, is outside the scope of this paper.

The differences between the spectroscopic and photometric effective
temperatures for the DB white dwarfs in our sample are displayed in
the bottom panel of Figure \ref{fig:deltaTeff}. Overall, the
photometric and spectroscopic temperatures are within the 1$\sigma$
confidence level for 69.6\% of the objects in our sample, as expected
from Gaussian statistics. As discussed in section \ref{sect:ivdw},
neutral broadening dominates below 16,000~K, and improvement in the
treatment of van der Waals broadening is required in our models. If
we exclude the objects cooler than this temperature, 71.8\% of our
sample is now within 1$\sigma$. Since the use of the Deridder \& van
Rensbergen theory rather than the Unsold theory has lowered the
effective temperatures obtained with the spectroscopic technique (see
top panel of Figure \ref{fig:ivdw_effect}), a more refined treatment
of van der Waals broadening will most likely lower those temperatures
even further, giving us a better agreement between $T_{\rm spec}$ and
$T_{\rm phot}$ in this regime.

We mentioned earlier that the SDSS spectroscopic data might still have
a calibration issue, and that this could be the cause of the
systematic offset in temperature ($T_{\rm spec}>T_{\rm phot}$)
observed in Figure \ref{fig:deltaTeff} for the DA stars. However,
residual calibration problems would also affect the DB spectroscopic
data. The fact that the offset in temperature is not seen in the DB
sample --- except for $T_{\rm spec}\lesssim 16,000$~K where we know
our spectroscopic effective temperatures are more uncertain ---
suggests that the root of the problem most likely lies within our
models, and not in the data.

Finally, the increased scatter at the high end of the temperature
distribution for the DB white dwarfs in Figure \ref{fig:deltaTeff} is
again due to the lack of sensitivity of the photometric technique, and
could possibly be reduced by combining multiple photometric systems.

\subsection{Stellar Masses}\label{sect:Mass}

\begin{figure*}[t!]
\includegraphics[width=\linewidth]{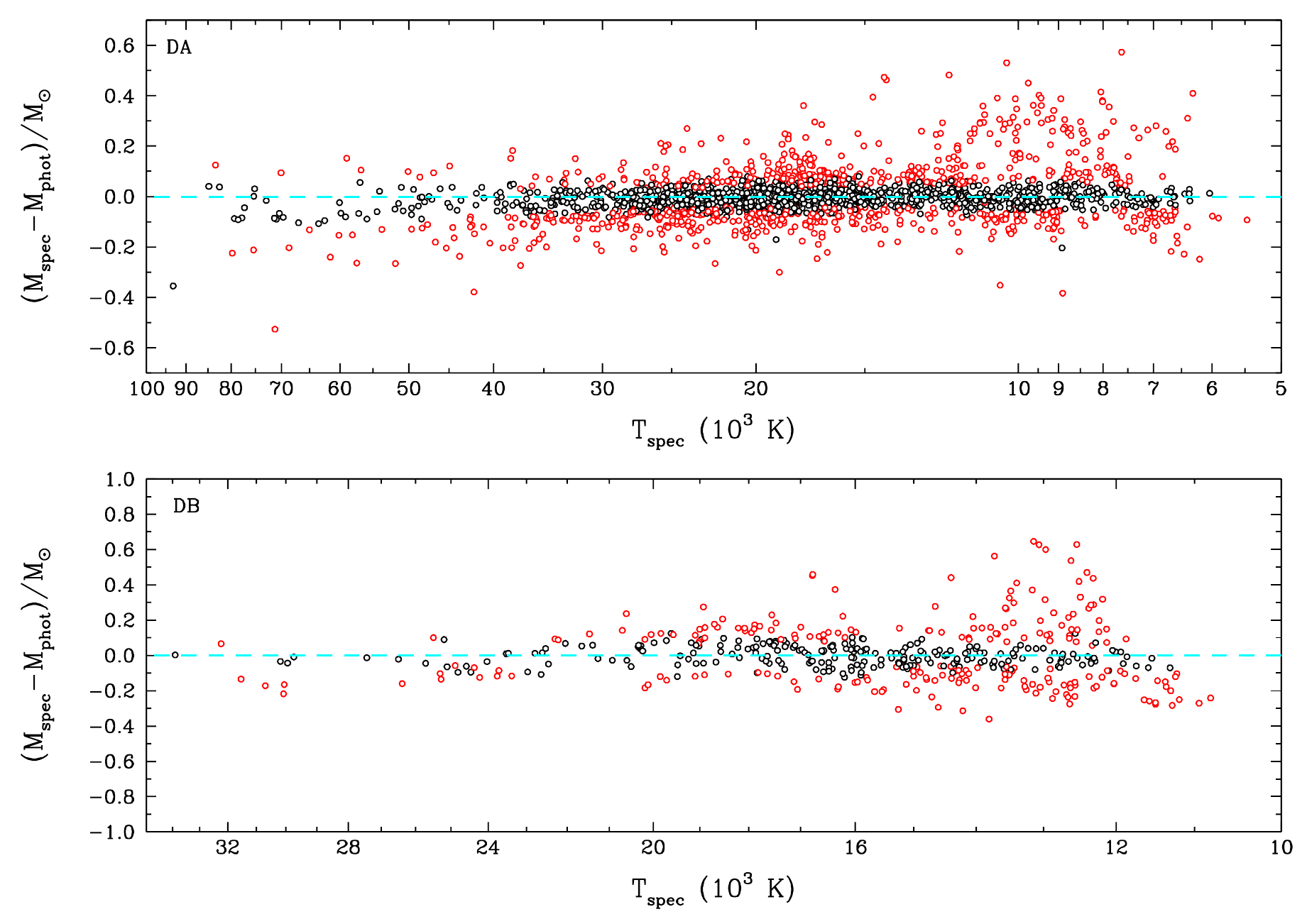}
\caption{Differences between spectroscopic and photometric masses as a
  function of $T_{\rm spec}$ for the DA (top) and DB (bottom) white
  dwarfs in our SDSS sample. Objects with mass estimates within
  1$\sigma$ are shown in black, and those outside are in red. The
  dashed line corresponds to $M_{\rm spec}=M_{\rm phot}$.}
\label{fig:deltaM}
\end{figure*}

The differences between the spectroscopic and photometric masses ---
$M_{\rm spec}$ and $M_{\rm phot}$ --- for the DA white dwarfs in our
sample are shown as a function of $T_{\rm spec}$ in the top panel of
Figure \ref{fig:deltaM}.  Unlike for the effective temperature, the
distribution of mass differences shows no obvious systematic effect,
which suggests that {\it the spectroscopic mass scale is more reliable
  than the corresponding temperature scale}, and less affected by the
problems with Stark broadening profiles discussed above.

\begin{figure}[t!]
        \includegraphics[width=\columnwidth]{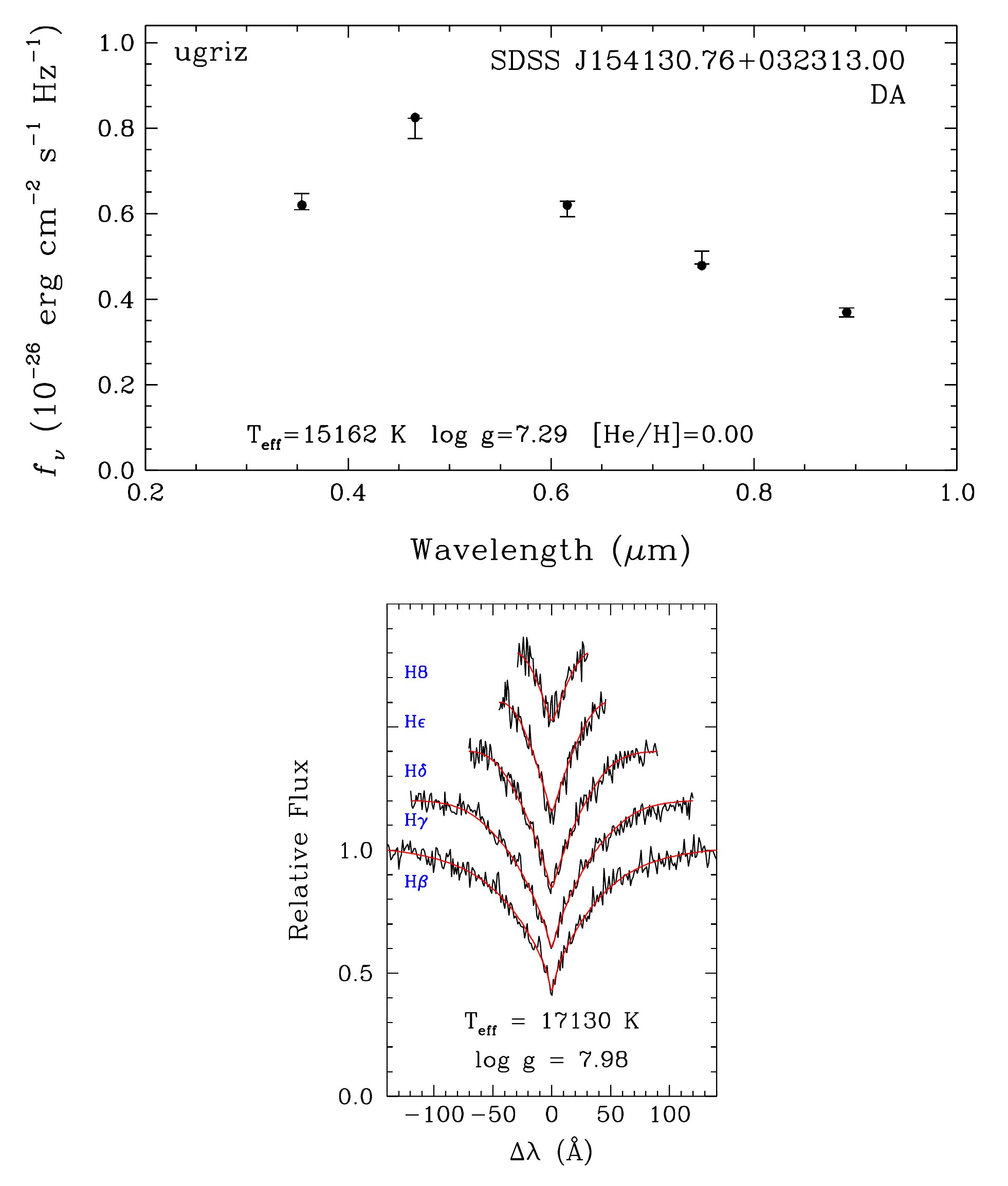}
        \caption{Best photometric (top) and spectroscopic (bottom)
fits to SDSS
  J154130.76+032313.00, a DA+DA double degenerate candidate.}
        \label{fig:DD_DA}
\end{figure}

Overall, the photometric and spectroscopic masses agree within the
1$\sigma$ confidence level for 60.9\% of the objects in our sample,
slightly below what is expected from Gaussian statistics.  However, we
can see in Figure \ref{fig:deltaM} a significant number of DA stars
with $M_{\rm spec}-M_{\rm phot} \gtrsim 0.2\ M_\odot$, which most
likely correspond to unresolved DA+DA double degenerate binaries. As
discussed in Section \ref{sect:photresults}, these objects have very
low inferred photometric masses, while the combined spectrum resembles
that of a single DA white dwarf with intermediate effective
temperature and mass \citep{Liebert1991}.  An example of such a double
degenerate candidate in our sample is SDSS J154130.76+032313.00, shown
in Figure \ref{fig:DD_DA}. The photometric fit indicates a very low
surface gravity for this object, $\log g = 7.29$, while the
spectroscopic solution yields a normal value of $\log g = 7.98$. Hence
this system is most likely composed of two normal mass DA stars. Note
that both the spectroscopic and photometric fits appear totally
normal. This means that DA+DA unresolved binaries can be difficult to
detect if one relies only on spectroscopic data.  If we omit the
binary candidates from our sample, we now find that 64\% of the DA
stars have photometric and spectroscopic masses that agree within
1$\sigma$, closer to what is expected from Gaussian statistics.

The mass comparison for the DB white dwarfs is presented in the bottom
panel of Figure \ref{fig:deltaM}. For this sample, only half of the
objects are within the 1$\sigma$ confidence level, significantly below
the Gaussian statistics. However, the number of DB stars outside
1$\sigma$ is dominated by objects at low temperatures, where van der
Waals broadening in our models yields uncertain spectroscopic
masses. If we restrict our sample to $T_{\rm eff} > 16,000$~K, the
proportion of objects within 1$\sigma$ increases to 60.5\%, much
closer to the expectation from Gaussian statistics. The second factor
that could improve the agreement between spectroscopic and photometric
masses is the use of 3D hydrodynamical models, which are expected to
affect the masses mostly between 20,000 K and 16,000 K, as discussed
in Section \ref{sect:specresults}. However, we note in Figure
\ref{fig:deltaM} that our mass estimates already agree fairly well in
this particular range of temperature, suggesting that 3D effects must
be small.

\begin{figure}
\includegraphics[width=\columnwidth]{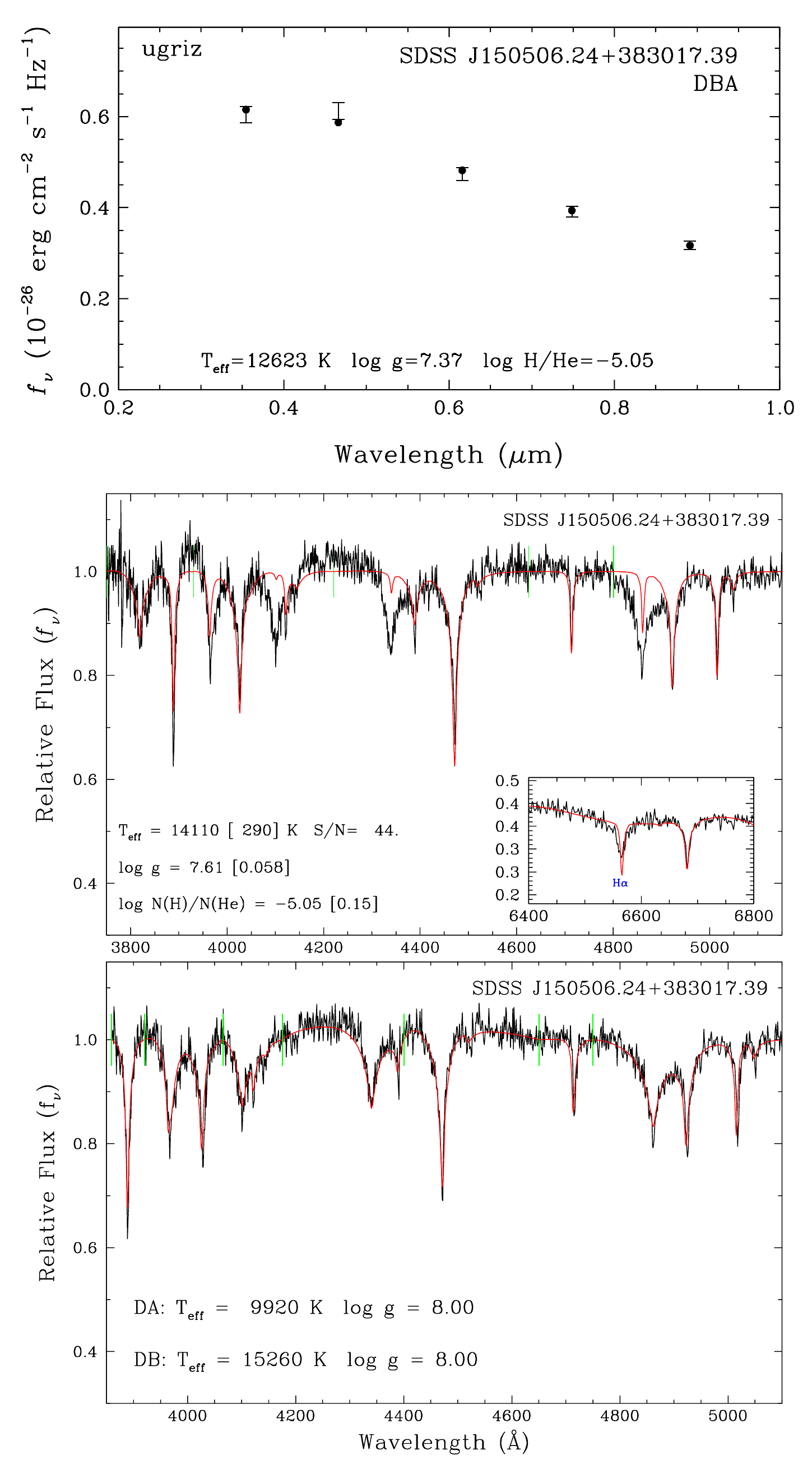}
\caption{Top panel: Best photometric fit to the DBA white dwarf SDSS
  J150506.24+383017.39.  Middle panel: Best spectroscopic fit of the
  same white dwarf under the assumption of a single star. Bottom
  panel: same as middle panel, but by assuming a DB+DA double
  degenerate binary.}
\label{fig:DB+DA}
\end{figure}

As for DA white dwarfs, our DB sample is likely to contain unresolved
double degenerate binaries, an example of which is the DBA white dwarf
SDSS J150506.24+383017.39, displayed in Figure \ref{fig:DB+DA}.  The
photometric fit indicates an extremely low surface gravity of $\log g
= 7.37$, or a mass of 0.302 $M_\odot$. Also, our spectroscopic fit for
the same object, under the assumption of a single star, reproduces the
hydrogen lines very poorly (middle panel of Figure
\ref{fig:DB+DA}). However, if we attempt to fit the same spectrum as a
combination of a DB and a DA white dwarf (here we assume $\log g=8.0$
for both components for simplicity), we are able to reproduce both the
hydrogen and helium lines perfectly, as shown in the bottom panel of
Figure \ref{fig:DB+DA}.

\begin{figure}
\includegraphics[width=\columnwidth]{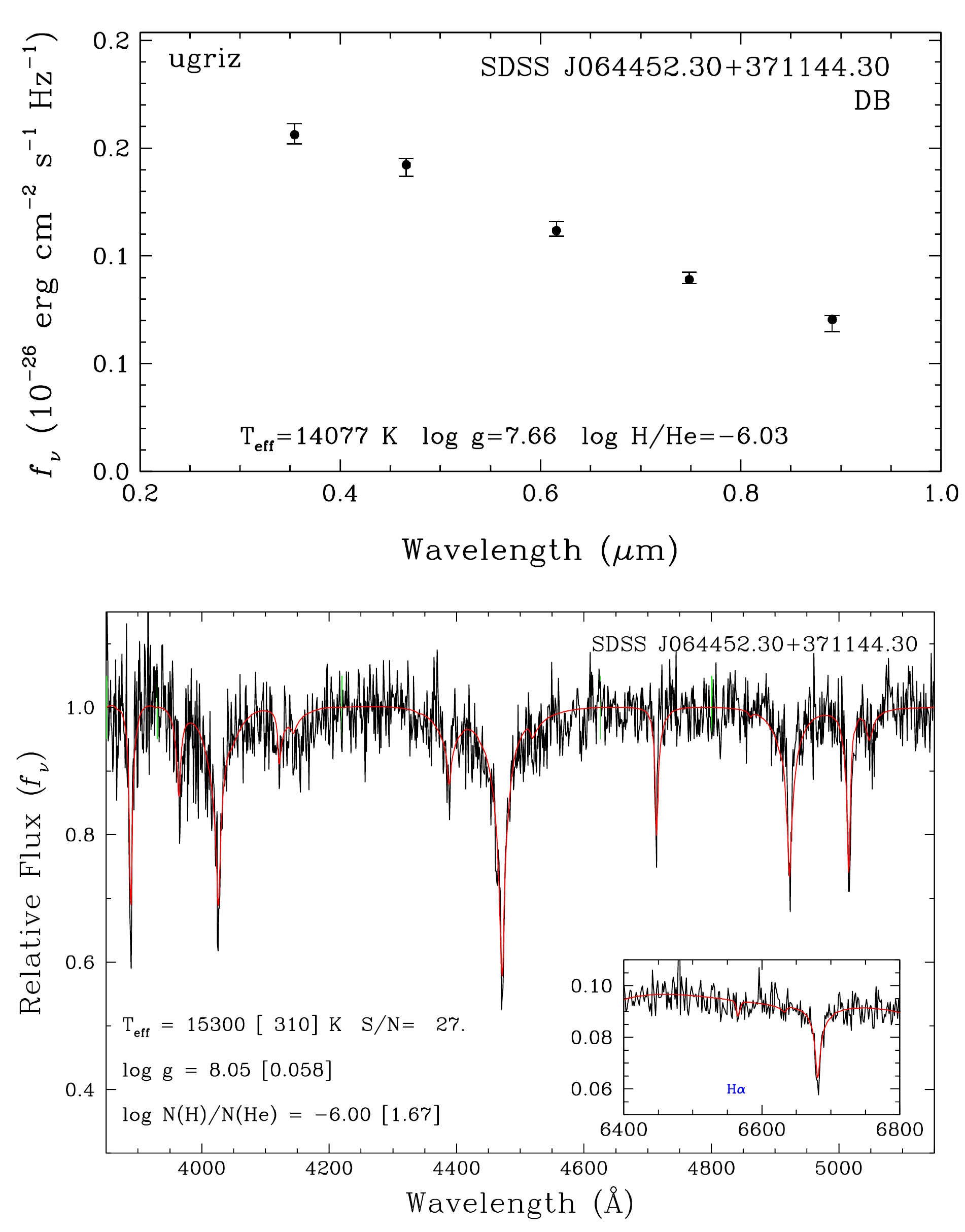}
\caption{Best photometric (top) and spectroscopic (bottom) fits to
  SDSS J064452.30+371144.30, an unresolved DB+DB double degenerate
  candidate.}
\label{fig:DD_DB}
\end{figure}

The DB white dwarf SDSS J064452.30+371144.30 is another example. Its
photometric fit, shown in the top panel of Figure \ref{fig:DD_DB},
indicates a surface gravity of $\log g = 7.66$, or a mass of
$M=0.412~M_\odot$.  In this case, however, we are able to successfully
reproduce the optical spectrum with single star models, as shown in
the bottom panel of Figure \ref{fig:DD_DB}. Furthermore, the
spectroscopic surface gravity has a normal value of $\log g =
7.96$. This suggests that this object is an unresolved double
degenerate binary composed of two DB stars with more normal masses,
where the combined spectrum resembles that of a single DB white dwarf
with intermediate atmospheric parameters. The deconvolution of this
system will be presented elsewhere.

\citet[][see also \citealt{Beauchamp1996}]{Bergeron2011} found no
evidence for the existence of low-mass ($M<0.5\ M_\odot$) DB white
dwarfs in their sample, suggesting that common envelope scenarios,
which are often invoked to explain low-mass DA white dwarfs, are not
producing DB stars. If we exclude all double degenerate binary
candidates from our DB sample, as well as all cool DB white dwarfs for
which spectroscopic masses are unreliable due our improper treatment
of van der Waals broadening, we find no evidence for low-mass DB white
dwarfs, in agreement with the conclusions of previous investigations.

\begin{figure}
\includegraphics[width=\columnwidth]{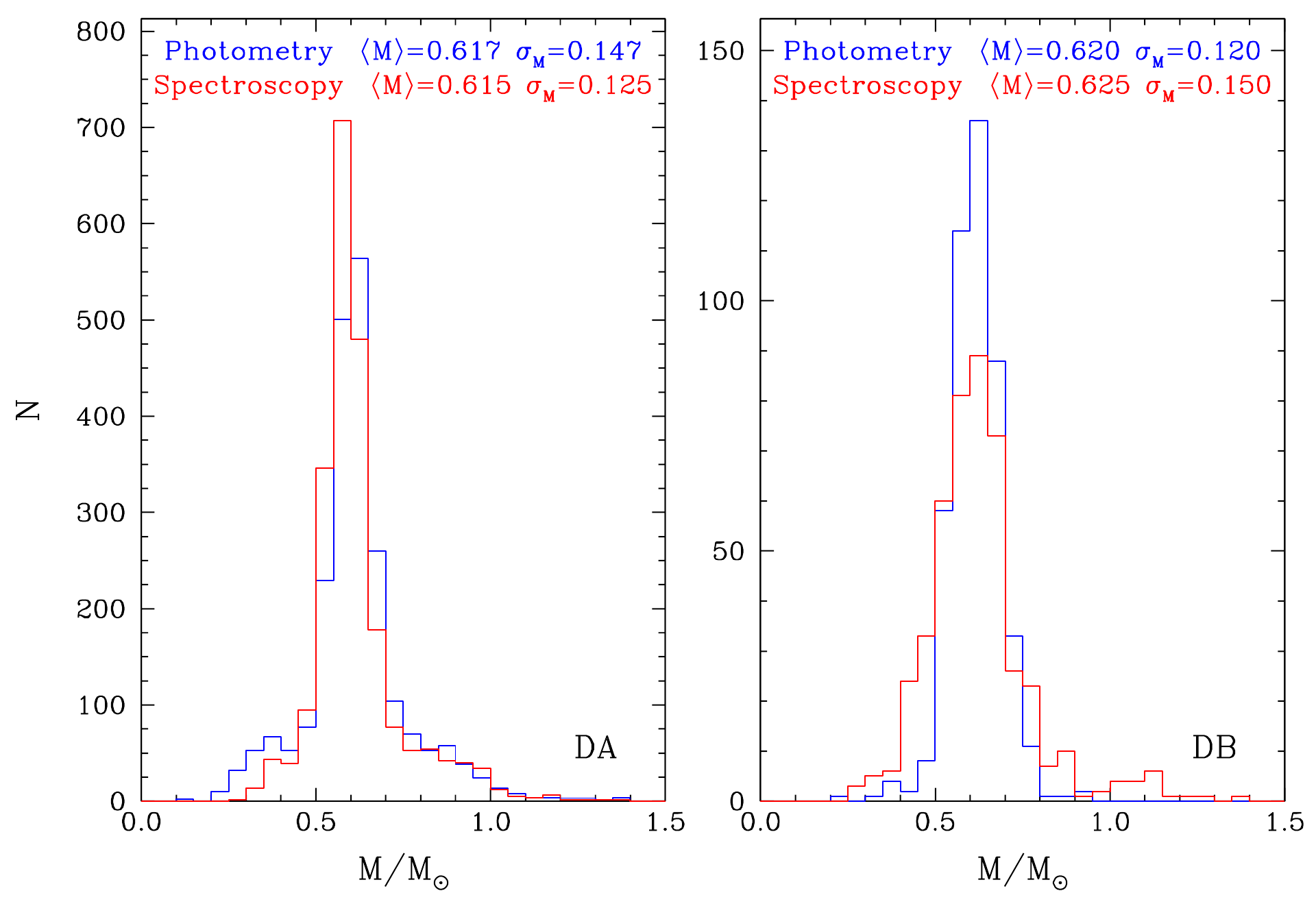}
\caption{Photometric (blue) and spectroscopic (red) cumulative mass
  distributions for the DA (left) and DB (right) white dwarfs in our
  sample. The corresponding mean masses and dispersions are given in
  each panel.}
\label{fig:histoM}
\end{figure}

To end our mass comparison, we present in Figure \ref{fig:histoM} the
cumulative photometric and spectroscopic mass distributions for both
the DA and DB white dwarfs in our sample. For the DA stars, the
photometric and spectroscopic distributions are in excellent
agreement, in particular the mean mass values, $\langle M_{\rm phot}
\rangle = 0.617~M_\odot$ and $\langle M_{\rm spec} \rangle =
0.615~M_\odot$.  These values are entirely consistent with previous
spectroscopic studies, for instance \cite{LBH05} using the PG sample
($\langle M \rangle=0.603~M_\odot$), \cite{Tremblay2011} using DA
white dwarfs from the SDSS DR4 ($\langle M \rangle=0.613~M_\odot$),
and \cite{Genest2014} using DA white dwarfs from the SDSS DR7
($\langle M \rangle=0.609~M_\odot$). The dispersion of the photometric
distribution ($\sigma_{\rm phot}=0.147~M_\odot$) is somewhat larger
than that of the spectroscopic distribution ($\sigma_{\rm
  spec}=0.125~M_\odot$), which is caused by the presence of unresolved
double degenerates in our sample.  These objects form the small bump
at $\sim 0.4~M_\odot$, and affect the photometric masses more
significantly.

In the case of DB stars, the photometric and spectroscopic mass
distributions have very different shapes (right panel of Figure
\ref{fig:histoM}) despite the fact that the mean masses, $\langle
M_{\rm phot} \rangle = 0.620~M_\odot$ and $\langle M_{\rm spec}
\rangle = 0.625~M_\odot$, are in excellent agreement. In particular,
the spectroscopic distribution shows both a low-mass and a high-mass
tails that are not observed in the photometric distribution. An
examination of Figure \ref{fig:MvsT_DB} (bottom panel) reveals that
all spectroscopic low-mass and high-mass DB stars in our sample are at
low temperatures, $T_{\rm eff} < 16,000~{\rm K}$, where the treatment
of van der Waals broadening is problematic. Hence both tails are
simple artifacts due to unreliable spectroscopic masses. As discussed
above, we find no evidence for low-mass DB stars in the photometric
mass distribution, with the exception of a small bump around 0.3
$M_\odot$ caused by the unresolved double degenerates in our sample.
 
Other mean mass values for DB white dwarfs, determined from
spectroscopy, are reported in the literature --- $\langle M \rangle =
0.596~M_\odot$ \citep[][based on SPY spectra]{Voss2007} and
$0.671~M_\odot$ \citep{Bergeron2011}. Similarly, \citet{Koester2015}
obtained a mean mass of $0.706~M_\odot$ for their complete sample of
DB stars from the SDSS DR10 and 12, but a lower value of $\langle M
\rangle = 0.606~M_\odot$ when their sample was restricted to
$16,000~{\rm K} \leq T_{\rm eff} \leq 22,000~{\rm K}$.  Note that
these mean mass values cannot be compared easily because of
differences in the treatment of van der Waals broadening in the
models, and even in the assumed convective efficiency.

Finally, we note that the mean masses for white dwarfs in the SDSS
obtained by \citet[][see their Figure 13, right panel]{Tremblay2019}
based on {\it Gaia} photometry --- $\langle M \rangle = 0.586~M_\odot$
and $0.580~M_\odot$ for DA and DB stars, respectively --- are 0.03
to $0.04~M_\odot$ smaller than our own photometric masses based on
$ugriz$ photometry. We will explore these differences in a future
publication.
 
\section{The Mass-Radius Relation for White Dwarfs}\label{sect:MR}

The photometric technique allows us to measure the effective
temperature $T_{\rm eff}$ and the solid angle $\pi(R/D)^2$,
and thus the stellar radius $R$ if the distance $D$ is known
from trigonometric parallax measurements. The
spectroscopic technique, on the other hand, yields $T_{\rm eff}$,
$\log g$, and the hydrogen abundance (or limits) in the case of DB white dwarfs.
In both cases, however, the mass of the object can only be obtained
from the theoretical mass-radius relation for white dwarfs, which
we put to the test in this section using the results obtained so far.

Since we have trigonometric parallax measurements, we know the precise
distance to every white dwarf in our sample. We can thus compare this
parallactic distance $D_\pi$ to the distance obtained from the
mass-radius relation, $D_{\rm MR}$, calculated using the procedure
outlined in \cite{Bedard2017}. First, the spectroscopic $\log g$ value
is converted into radius $R$ using evolutionary models, which is then
combined with the photometric solid angle $\pi(R/D)^2$ to obtain the
desired distance $D_{\rm MR}$. When using this approach, the solid
angle $\pi(R/D)^2$ can be measured in two different ways. The first
one is to consider both the effective temperature and the solid angle
free parameters in the minimization procedure; the second is to force
the effective temperature to the spectroscopic value and to fit only
the solid angle.  Given the uncertainties with the spectroscopic
temperature scales discussed in Section \ref{sect:comparison}, we
adopt the former approach. Also, it is justified to rely on
spectroscopic $\log g$ values for this exercise since the surface
gravity (or mass) scale appears reasonably accurate according to the
results shown in Figure \ref{fig:deltaM}.

\begin{figure*}
\includegraphics[width=\linewidth]{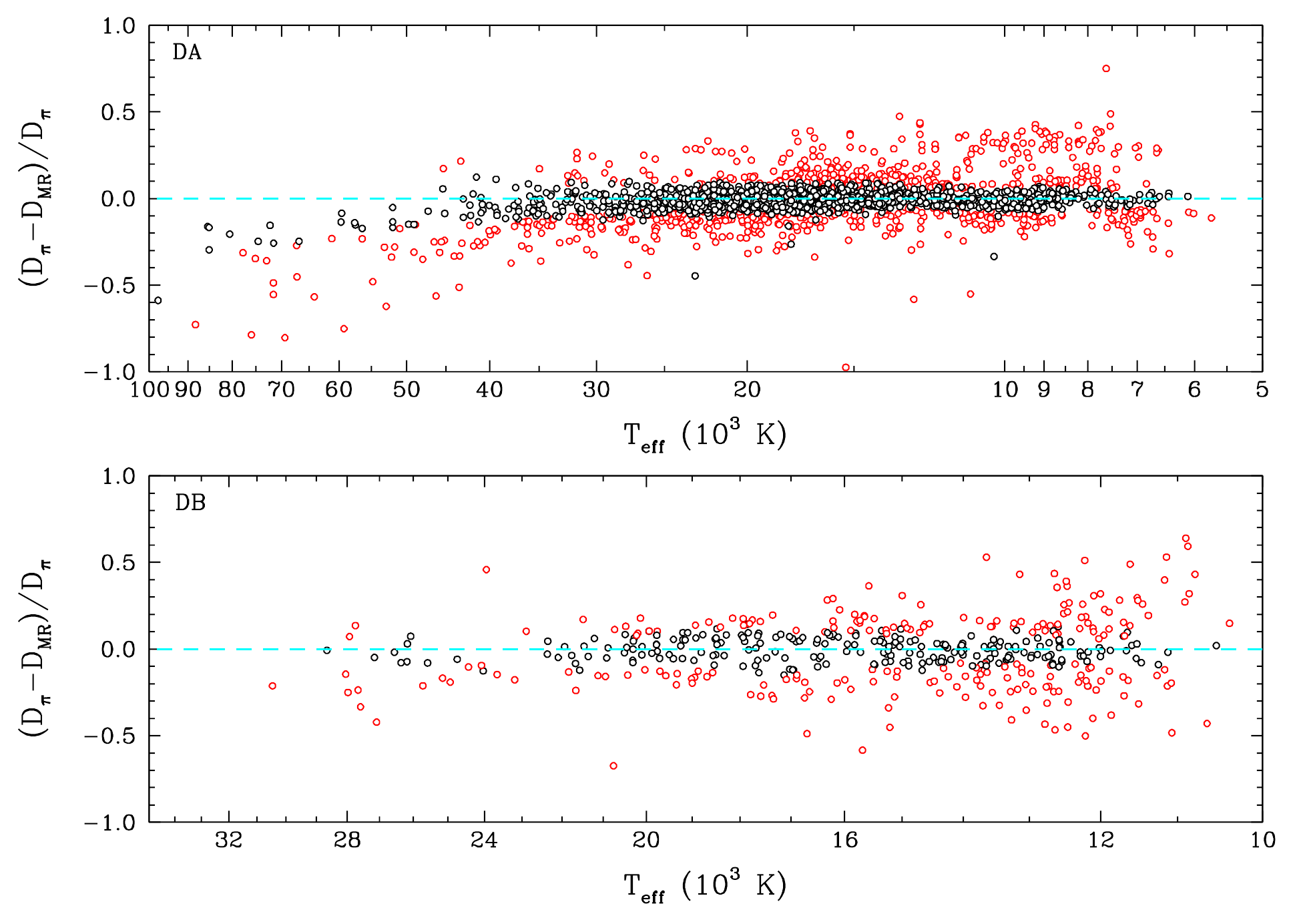}
\caption{Comparison between the parallactic distance $D_\pi$ and
  the distance obtained from the mass-radius relation $D_{\rm MR}$, as
  a function of effective temperature, for both the DA (top) and DB
  (bottom) white dwarfs in our sample. The objects for which the
  distance estimates are within the 1$\sigma$ confidence level are
  shown in black.}
\label{fig:DMR}
\end{figure*}

Figure \ref{fig:DMR} shows the difference between $D_\pi$ and
$D_{\rm MR}$ as a function of effective temperature, for both the DA
and DB white dwarfs in our sample.  For the DA stars, these distance
estimates are within the 1$\sigma$ confidence level for $61.4\%$ of
the objects in our sample, a value somewhat lower than what is
expected from Gaussian statistics. However, we can see from this
figure that most of the outliers are found (1) at high temperatures
($T_{\rm eff} \gtrsim 40,000~{\rm K}$) where the energy distribution
sampled by the $ugriz$ photometry is in the Rayleigh-Jeans regime, and
(2) at the top of the figure where unresolved double degenerate binaries
are expected. If we drop all the objects above 40,000~K, as well as all
the binary candidates, the fraction of objects within 1$\sigma$
increases to $64.6\%$, again much closer to the expected value of $68\%$.

For the DB stars in Figure \ref{fig:DMR}, we find only 49.2\% of the
objects in our sample with distance estimates that are within the
1$\sigma$ confidence level, a value significantly lower than the
expected 68\%. Here we see, however, that most of the outliers are
located at low effective temperatures where van der Waals broadening
becomes important. If we restrict our sample to $T_{\rm eff} >
16,000~{\rm K}$, and also omit the double degenerate binary
candidates, the fraction of objects within 1$\sigma$ increases to
$58.5\%$. This is still short of the expected fraction, but the number
of objects left in our sample if we exclude the cool DB stars is
admittedly small.

\begin{figure}
\includegraphics[width=\columnwidth]{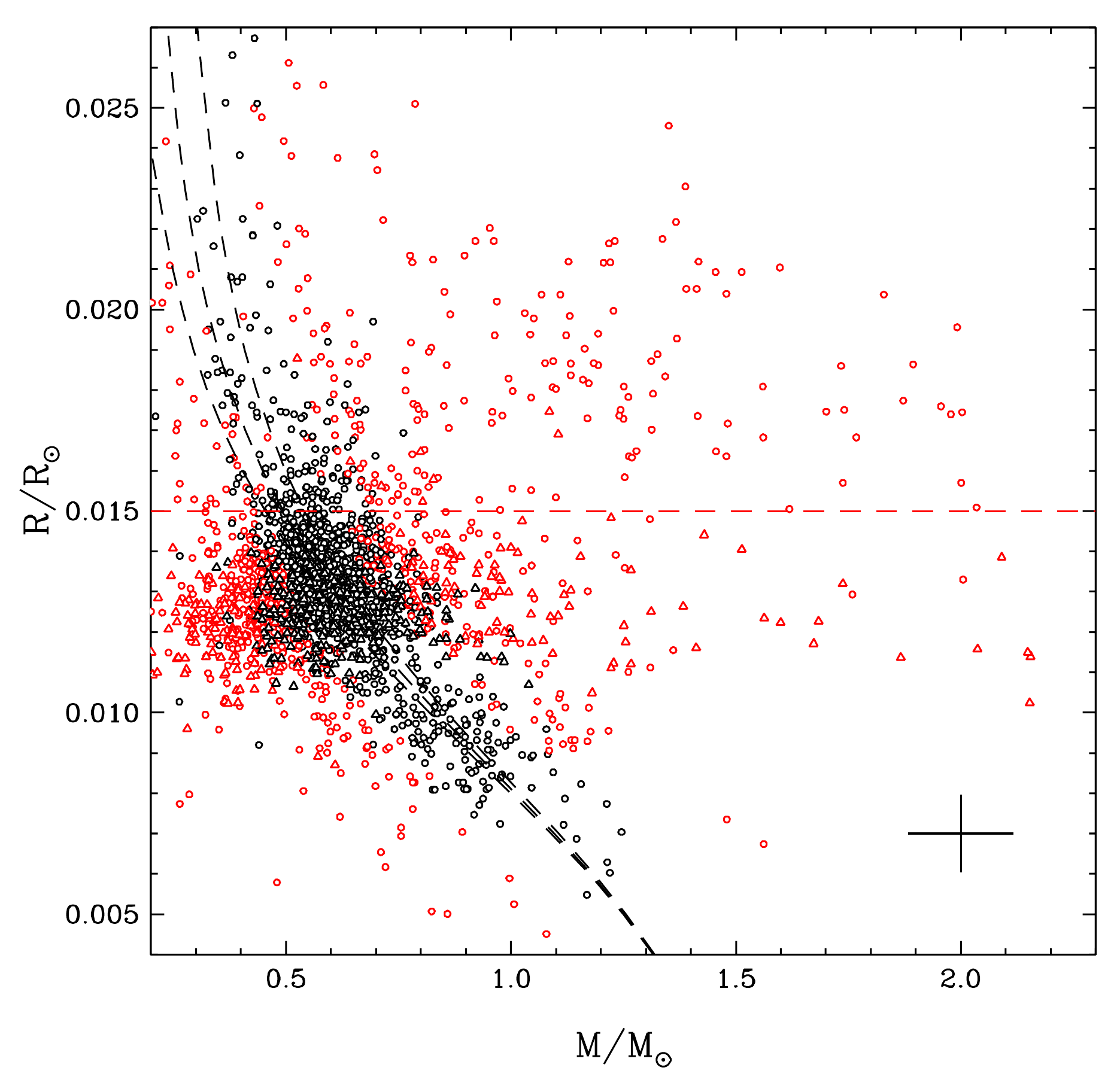}
\caption{Radius as a function of mass for all the DA (circles) and DB
  (triangles) white dwarfs in our sample. The cross in the lower right
  corner represents the average uncertainties. Objects for which the
  difference between $D_\pi$ and $D_{\rm MR}$ exceeds the 1$\sigma$
  confidence level are shown in red. Also shown are the theoretical
  mass-radius relations for C/O-core, thick hydrogen envelope models
  at $T_{\rm eff}=8000$, 15,000, and 25,000~K (black dashed lines,
  from left to right). The dashed red horizontal line is located at
  $R= 0.015~R_\odot$ (see text).}
\label{fig:MRrelation}
\end{figure}

Another way to test the mass-radius relation is to plot the stellar
radius $R$, obtained {\it directly} from the photometric technique,
against the mass obtained by combining this photometric radius with
the spectroscopic $\log g$ ($g=GM/R^2$). This procedure allows us to
measure the radius and the mass of an object without the use of any
theoretical mass-radius relation. Our results for all the DA and DB
white dwarfs in our sample are displayed in Figure
\ref{fig:MRrelation}, together with theoretical mass-radius relations
(see Section \ref{sect:techphot}) for C/O-core, thick hydrogen
envelope models at various representative $T_{\rm eff}$ values. Note
that in such a diagram, thin layer models would be almost impossible
to distinguish from thick layer models (see \citealt{Bedard2017}).  As
in B\'edard et al., we use different color symbols to indicate objects
that exhibit differences larger than the 1$\sigma$ confidence level
between the two distance estimates $D_\pi$ and $D_{\rm MR}$ introduced
above.

While most of the data points (the black symbols) align well on the
expected mass-radius relation, we see a very large scatter, especially
towards higher masses. As discussed in \citet{Bedard2017}, the upper
right corner of this diagram is populated by unresolved double
degenerate binaries. For such overluminous systems, the stellar radius
is overestimated, while the spectroscopic $\log g$ value appears
normal, resulting into a large inferred mass in the mass-radius
diagram. For example, for the DA+DA binary candidate SDSS
J154130.76+032313.00 displayed in Figure \ref{fig:DD_DA}, we obtain
from the photometric technique $R=0.02093~R_\odot$ and a spectroscopic
value of $\log g_{\rm spec} = 7.98$, resulting in a large mass of
$M=1.513~M_\odot$, well above the Chandrasekhar limit. 

Single star evolution predicts that white dwarfs with masses below
$M\sim 0.45\ M_\odot$ should not have formed within the lifetime of
the Galaxy. This corresponds to a stellar radius of $\sim$0.015
$R_\odot$ at 10,000 K, represented by the horizontal dashed line in
Figure \ref{fig:MRrelation}. We see that the most massive objects
below this line --- with normal photometric radii around 0.012
$R_\odot$ --- are DB white dwarfs. More specifically, these correspond
to the cool DB stars in our sample with $T_{\rm eff} < 16,000~{\rm
  K}$, and with uncertain high $\log g$ values (see bottom panel of
Figure \ref{fig:ivdw}). Similarly, we see a large concentration of
low-mass DB white dwarfs (red triangles) on the extreme left around
0.012 $R_\odot$. These correspond to DB stars in the same temperature
range, but this time with low spectroscopic $\log g$ values (see again
bottom panel of Figure \ref{fig:ivdw}). All of these cool DB white
dwarfs are expected to align on the mass-radius relation once a proper
treatment of van der Waals broadening becomes available.

If we remove the DB stars from our sample, we are still left with a
fairly large number of DA stars to the left of the theoretical curves
with discrepant ($>$1$\sigma$) distance estimates (red circles).
\cite{Provencal1998} proposed that these white dwarfs might not have a
C/O core, but rather an iron core. \cite{Bedard2017} tested this
hypothesis and found that Fe-core models could indeed explain the
location of these objects in the mass-radius diagram, but they could
not completely rule out that these were not normal C/O-core white
dwarfs, within the uncertainties. The most compelling case of an
Fe-core white dwarf was G87-7, with a parallactic distance of
$D_{\pi}=15.7$ pc based on {\it Hipparcos}. The distance obtained from
C/O-core models, $D_{\rm MR}=17.5$ pc, was found to be significantly
different from the parallactic distance, while the distance inferred
from Fe-core models, $D_{\rm MR}=15.9$ pc, was in much better
agreement. This discrepancy has now been resolved with the {\it Gaia}
parallax, which yields $D_\pi=17.07$ pc, in excellent agreement with
the distance inferred from C/O-core models.

To summarize, if we restrict our analysis to single white dwarfs
($M_{\rm spec}-M_{\rm phot} < 0.2\ M_\odot)$, and to the temperature
range where the physics of our model atmospheres is better understood
($T_{\rm eff} < 40,000~{\rm K}$ for DA stars, and $T_{\rm eff} >
16,000~{\rm K}$ for DB stars), we find that about 65\% are within the
1$\sigma$ confidence level of the mass-radius relation (based on the
distance comparison), and about 92\% are within 2$\sigma$. We stress
again the fact that we are probably underestimating in our analysis
the uncertainties associated with the \textit{spectroscopic}
atmospheric parameters, thus these numbers represent only lower
limits. With these caveats in mind, we conclude that the theoretical
mass-radius relation for white dwarfs rests on solid empirical
grounds, a conclusion also reached by \citet{Holberg12},
\citet{Tremblay17}, \citet{Parsons17}, and \cite{Bedard2017}, but for
significantly smaller samples.

\section{Conclusion}\label{sect:conclusion}

We performed a detailed spectroscopic and photometric analysis of 2236
DA and 461 DB white dwarfs from the Sloan Digital Sky Survey with
trigonometric parallax measurements available from the {\it Gaia}
mission. The temperature and mass scales obtained from fits to $ugriz$
photometry appear reasonable for both DA and DB stars.  The
photometric mass distributions for DA and DB stars are comparable,
with almost identical mean masses of $\langle M \rangle =
0.617~M_\odot$ and $0.620~M_\odot$, respectively. However, the DA mass
distribution shows well-defined low-mass and high-mass tails, which are not
observed in the DB photometric mass distribution. In particular, we find
no evidence in our sample for single, low-mass DB white dwarfs.

The comparison of the effective temperatures and stellar masses
obtained from the photometric and spectroscopic techniques reveals
several problems with the model spectra for both pure hydrogen and
pure helium compositions. For DA stars, we found a systematic offset
in temperature, with the spectroscopic temperatures exceeding the
photometric values by $\sim$10\% above 14,000 K. Since this offset is
not observed for DB stars in the same temperature range, we believe
that some inaccuracy in the theory of Stark broadening for hydrogen lines
is at the origin of the observed temperature discrepancies. Despite these
problems, the $\log g$ and mass scales derived from spectroscopy
appear unaffected since the spectroscopic mass distribution agrees
extremely well with that obtained from photometry. For instance, the
spectroscopic mean mass for the DA stars, $\langle M \rangle =
0.615~M_\odot$, differs from the photometric mean value by only
$0.002~M_\odot$.

For the DB white dwarfs, both the temperature and mass scales agree
well above 16,000~K, but abnormally low and high spectroscopic masses
are found at lower temperatures that are significantly different
from the corresponding photometric masses. We attribute these
discrepancies to the inaccurate treatment of van der Waals broadening
in our model spectra for DB white dwarfs, as well as to the
limitations of the spectroscopic technique at low effective
temperatures, when the neutral helium lines become too weak. Despite
these problems, the spectroscopic mean mass for the DB stars, $\langle
M \rangle = 0.625~M_\odot$, differs by only $0.005~M_\odot$ from the
photometric mean value.

By comparing the physical parameters using both the photometric and
spectroscopic techniques, we were able to identify very easily several
unresolved double degenerate binaries in our sample with various
spectral types, including DA+DA, DB+DB, DA+DB, and even DA+DC systems.
All of these appear overluminous, and thus have extremely low
photometric masses. Double degenerates composed of identical spectral
types may have normal spectroscopic $\log g$ values, however.

We finally took advantage of the {\it Gaia} parallaxes to test the
theoretical mass-radius relation for white dwarfs. If we exclude the
double degenerate binary candidates from our sample, and restrict our
analysis to the temperature range where the spectroscopic $\log g$
values are reasonably accurate, we find that the parallactic distance
and the distance obtained from the mass-radius relation are within the
1$\sigma$ confidence level for about 65\% of the white dwarfs in our
sample, which confirms the validity of the theoretical mass-radius
relation for white dwarfs.

\acknowledgments We are grateful to A. B\'edard for a careful reading
of our manuscript, and to the referee, D. Koester, for his
constructive comments and suggestions. We would also like to thank
P.-E.~Tremblay and E.~Cukanovaite for useful discussions, and in
particular for mentioning to us the problem with our former use of van
der Waals broadening. This work is supported in part by the NSERC
Canada and by the Fund FRQ-NT (Qu\'ebec). This work has made use of
data from the European Space Agency (ESA) mission {\it Gaia}
(\url{https://www.cosmos.esa.int/gaia}), processed by the {\it Gaia}
Data Processing and Analysis Consortium (DPAC,
\url{https://www.cosmos.esa.int/web/gaia/dpac/consortium}). Funding
for the DPAC has been provided by national institutions, in particular
the institutions participating in the {\it Gaia} Multilateral
Agreement. This research has made use of the NASA/ IPAC Infrared
Science Archive, which is operated by the Jet Propulsion Laboratory,
California Institute of Technology, under contract with the National
Aeronautics and Space Administration.

\bibliographystyle{aasjournal}
\bibliography{references}

\end{document}